\documentclass[10pt]{article}
\usepackage{amssymb}

\usepackage{graphicx}
\usepackage{amsmath}

\input{tcilatex}

\begin{document}

\centerline{\textbf{\Large}}

\vskip 0.8truecm

\centerline{\textbf{\Large VARIATIONAL TWO FERMION WAVE EQUATIONS}} \vskip %
0.3truecm \centerline{\textbf{\Large IN QED: MUONIUM LIKE SYSTEMS}}

\vskip 0.6truecm 
\centerline{\large Andrei G. Terekidi$^{\dagger }$ and
Jurij W. Darewych$^{\dagger \dagger }$}

\vskip 0.5truecm 
\centerline{\footnotesize \emph{Department of Physics and
Astronomy, York University, Toronto, Ontario, M3J 1P3, Canada}} 
\centerline{\footnotesize \emph{$^{\dagger }$terekidi@yorku.ca, $^{\dagger
\dagger }$darewych@yorku.ca}}

\vskip 0.6truecm


\noindent \textbf{Abstract} We consider a reformulation of QED in which
covariant Green functions are used to solve for the electromagnetic field in
terms of the fermion fields. The resulting modified Hamiltonian contains the
photon propagator directly. A simple Fock-state variational trial function
is used to derive relativistic two-fermion equations variationally from the
expectation value of the Hamiltonian of the field theory. The interaction
kernel of the equation is shown to be, in essence, the invariant $\mathcal{%
M\,}\ $matrix in lowest order. Solutions of the two-body equations are
presented for muonium like system for small coupling strengths. The results
compare well with the observed muonium spectrum, as well as that for
hydrogen and muonic hydrogen. Anomalous magnetic moment effects are
discussed.

\vskip 0.8truecm

\noindent\textbf{\large 1. Introduction} 

\vskip0.4truecm It has been pointed out in previous publications that
various models in Quantum Field Theory (QFT), including QED, can be
reformulated, using mediating-field Green functions, into a form that is
particularly convenient for variational calculation [1,2]. This approach was
applied to the study of relativistic two-body eigenstates in the scalar
Yukawa (Wick-Cutkosky) theory [3,4,5]. We shall implement such an approach
to two-fermion states in QED in this paper.

The Lagrangian of two fermion fields interacting electromagnetically is ($%
\hbar =c=1$){\normalsize 
\begin{eqnarray}
\mathcal{L} &=&\overline{\psi }(x)\left( i\gamma ^{\mu }\partial _{\mu
}-m_{1}-q_{1}\gamma ^{\mu }A_{\mu }(x)\right) \psi (x)+\overline{\phi }%
(x)\left( i\gamma ^{\mu }\partial _{\mu }-m_{2}-q_{2}\gamma ^{\mu }A_{\mu
}(x)\right) \phi (x)  \notag \\
&&\;\;\;\;\;\;\;\;\;\;\;\;\;-\frac{1}{4}\left( \partial _{\alpha }A_{\beta
}(x)-\partial _{\beta }A_{\alpha }(x)\right) \left( \partial ^{\alpha
}A^{\beta }(x)-\partial ^{\beta }A^{\alpha }(x)\right) .
\end{eqnarray}
}The corresponding Euler-Lagrange equations of motion are the coupled
Dirac-Maxwell equations, {\normalsize 
\begin{equation}
\left( i\gamma ^{\mu }\partial _{\mu }-m_{1}\right) \psi (x)=q_{1}\gamma
^{\mu }A_{\mu }(x)\psi (x),
\end{equation}
} 
\begin{equation}
\left( i\gamma ^{\mu }\partial _{\mu }-m_{2}\right) \phi (x)=q_{2}\gamma
^{\mu }A_{\mu }(x)\phi (x),
\end{equation}
and 
\begin{equation}
\partial _{\mu }\partial ^{\mu }A^{\nu }(x)-\partial ^{\nu }\partial _{\mu
}A^{\mu }(x)=j^{\nu }(x),
\end{equation}
where 
\begin{equation}
j^{\nu }(x)=q_{1}\overline{\psi }(x)\gamma ^{\nu }\psi (x)+q_{2}\overline{%
\phi }(x)\gamma ^{\nu }\phi (x).
\end{equation}
Equations (2)-(3) can be decoupled in part by using the well-known formal
solution [6,7] of the Maxwell equation (4), namely 
\begin{equation}
A_{\mu }(x)=A_{\mu }^{0}(x)+\int d^{4}x^{\prime }D_{\mu \nu }(x-x^{\prime
})j^{\nu }(x^{\prime }),
\end{equation}
where $D_{\mu \nu }(x-x^{\prime })$ is a Green function (or photon
propagator in QFT terminology), defined by 
\begin{equation}
\partial _{\alpha }\partial ^{\alpha }D_{\mu \nu }(x-x^{\prime })-\partial
_{\mu }\partial ^{\alpha }D_{\alpha \nu }(x-x^{\prime })=g_{\mu \nu }\delta
^{4}(x-x^{\prime }),
\end{equation}
and $A_{\mu }^{0}(x)$ is a solution of the homogeneous (or ``free field'')
equation (4) with $j^{\mu }(x)=0.$

We recall that equation (7) does not define the covariant Green function $%
D_{\mu \nu }(x-x^{\prime })$ uniquely. For one thing, one can always add a
solution of the homogeneous equation (eq. (7) with $g_{\mu \nu }\rightarrow
0 $). This allows for a certain freedom in the choice of $D_{\mu \nu }$, as
is discussed in standard texts (e.g. ref. [6,7]). In practice, the solution
of eq. (7), like that \ of eq. (4), requires a choice of gauge. However, we
do not need to specify one at this stage.

Substitution of the formal solution (6) into equations (2) and (3) yields
the ``partly reduced'' equations,

{\normalsize 
\begin{eqnarray}
\left( i\gamma ^{\mu }\partial _{\mu }-m_{1}\right) \psi (x) &=&q_{1}\gamma
^{\mu }\left( A_{\mu }^{0}(x)+\int d^{4}x^{\prime }D_{\mu \nu }(x-x^{\prime
})j^{\nu }(x^{\prime })\right) \psi (x), \\
\left( i\gamma ^{\mu }\partial _{\mu }-m_{2}\right) \phi (x) &=&q_{2}\gamma
^{\mu }\left( A_{\mu }^{0}(x)+\int d^{4}x^{\prime }D_{\mu \nu }(x-x^{\prime
})j^{\nu }(x^{\prime })\right) \phi (x).
\end{eqnarray}
}These are coupled nonlinear Dirac equations. To our knowledge no exact
(analytic or numeric) solution of equations (8)-(9) for classical fields
have been reported in the literature, though approximate (perturbative)
solutions have been discussed by various authors, particularly Barut and his
co-workers (see ref. [8,9] and citations there in). However, our interest
here is in the quantized field theory.

The partially reduced equations (8)-(9) are derivable from the stationary
action principle 
\begin{equation}
\delta S\left[ \psi \right] =\delta \int d^{4}x\mathcal{L}_{R}=0
\end{equation}
with the Lagrangian density {\normalsize 
\begin{eqnarray}
\mathcal{L}_{R} &=&\overline{\psi }(x)\left( i\gamma ^{\mu }\partial _{\mu
}-m_{1}-q_{1}\gamma _{\mu }A_{0}^{\mu }(x)\right) \psi (x)+\overline{\phi }%
(x)\left( i\gamma ^{\mu }\partial _{\mu }-m_{2}-q_{2}\gamma _{\mu
}A_{0}^{\mu }(x)\right) \phi (x)  \notag \\
&&\;\;\;\;\;\;\;\;\;\;\;\;\;\;\;\;\;\;\;-\frac{1}{2}\int d^{4}x^{\prime
}j^{\mu }(x^{\prime })D_{\mu \nu }(x-x^{\prime })j^{\nu }(x)
\end{eqnarray}
}provided that the Green function is symmetric in the sense that 
\begin{equation}
D_{\mu \nu }(x-x^{\prime })=D_{\mu \nu }(x^{\prime }-x)\;\;\;\;\;and\;\;\
\;\ D_{\mu \nu }(x-x^{\prime })=D_{\nu \mu }(x-x^{\prime }).
\end{equation}
The interaction part of (11) has a somewhat modified structure from that of
the usual formulation of QED. Thus, there are two interaction terms. The
last term of (11) is a ``current-current'' interaction which contains the
photon propagator sandwiched between the fermionic currents. We shall use
this modified formulation together with a variational approach to obtain
relativistic few-fermion equations, and to study their bound state solutions.

\vskip0.8truecm {\normalsize \noindent {\textbf{\large 2. Hamiltonian }}%
\vskip0.4truecm }

We shall consider the quantized theory in the equal-time formalism. To this
end we write down the Hamiltonian density corresponding to the Lagrangian
(11), namely 
\begin{equation}
\mathcal{H}_{R}=\mathcal{H}_{0}\mathcal{+H}_{I}+\mathcal{H}_{II},
\end{equation}
where

\begin{eqnarray}
\mathcal{H}_{0} &=&\psi ^{\dagger }(x)\left( -i\overrightarrow{\alpha }\cdot
\nabla +m_{1}\beta \right) \psi (x)+\phi ^{\dagger }(x)\left( -i%
\overrightarrow{\alpha }\cdot \nabla +m_{2}\beta \right) \phi (x), \\
\mathcal{H}_{I} &=&\frac{1}{2}\int d^{4}x^{\prime }j^{\mu }(x^{\prime
})D_{\mu \nu }(x-x^{\prime })j^{\nu }(x), \\
\mathcal{H}_{II} &=&q_{1}\overline{\psi }(x)\,\gamma _{\mu }\,A_{0}^{\mu
}(x)\,\psi (x)+q_{2}\overline{\phi }(x)\,\gamma _{\mu }\,A_{0}^{\mu
}(x)\,\phi (x),
\end{eqnarray}
and where we have suppressed the kinetic-energy term of the free photon
field. We construct a quantized theory by imposing equal-time
anticommutation rules for the fermion fields, namely 
\begin{equation*}
\left\{ \psi _{\alpha }(\mathbf{x},t),\psi _{\beta }^{\dagger }(\mathbf{y}%
,t)\right\} =\left\{ \phi _{\alpha }(\mathbf{x},t),\phi _{\beta }^{\dagger }(%
\mathbf{y},t)\right\} =\delta _{\alpha \beta }\delta ^{3}\left( \mathbf{x}-%
\mathbf{y}\right) ,
\end{equation*}
and all other vanish. In addition, there are the usual commutation rules for
the $A_{0}^{\mu }$ field, and commutation of the $A_{0}^{\mu }$ field
operators with the $\psi $ and $\phi $ field operators.

To specify our notation, we quote the Fourier decomposition of the field
operators, namely 
\begin{equation}
\psi (x)=\sum_{s}\int \frac{d^{3}\mathbf{p}}{\left( 2\pi \right) ^{3/2}}%
\left( \frac{m_{1}}{\omega _{p}}\right) ^{1/2}\left[ b_{\mathbf{p}s}u\left( 
\mathbf{p},s\right) e^{-ip\cdot x}+d_{\mathbf{p}s}^{\dagger }v\left( \mathbf{%
p},s\right) e^{ip\cdot x}\right] ,
\end{equation}
with $p=p_{m}=\left( \omega _{p},\mathbf{p}\right) $, $\omega _{p}=\sqrt{%
m_{1}^{2}+\mathbf{p}^{2}}$ and

\begin{equation}
\phi (x)=\sum_{s}\int \frac{d^{3}\mathbf{p}}{\left( 2\pi \right) ^{3/2}}%
\left( \frac{m_{2}}{\Omega _{p}}\right) ^{1/2}\left[ B_{\mathbf{p}s}U\left( 
\mathbf{p},s\right) e^{-ip\cdot x}+D_{\mathbf{p}s}^{\dagger }V\left( \mathbf{%
p},s\right) e^{ip\cdot x}\right] ,
\end{equation}
with $p=p_{m}=\left( \Omega _{p},\mathbf{p}\right) $, $\Omega _{p}=\sqrt{%
m_{2}^{2}+\mathbf{p}^{2}}$. Note that the mass-$m_{1}$ free-particle Dirac
spinors $u\left( \mathbf{p},s\right) $, $v\left( \mathbf{p},s\right) $\
where $\left( \gamma ^{\mu }p_{\mu }-m_{1}\right) u\left( \mathbf{p}%
,s\right) =0$,\ $\left( \gamma ^{\mu }p_{\mu }+m_{1}\right) v\left( \mathbf{p%
},s\right) =0$, are normalized such that 
\begin{eqnarray}
u^{\dagger }\left( \mathbf{p},s\right) u\left( \mathbf{p},\sigma \right)
&=&v^{\dagger }\left( \mathbf{p},s\right) v\left( \mathbf{p},\sigma \right) =%
\frac{\omega _{p}}{m_{1}}\delta _{s\sigma }, \\
u^{\dagger }\left( \mathbf{p},s\right) v\left( \mathbf{p},\sigma \right)
&=&v^{\dagger }\left( \mathbf{p},s\right) u\left( \mathbf{p},\sigma \right)
=0.
\end{eqnarray}
Analogous properties apply to the mass-$m_{2}$ spinors $U$, $V$. The
creation and destruction operators $b^{\dagger }$, $b$ of the (free)
particles of mass $m_{1}$, and $d^{\dagger }$, $d$ for the corresponding
antiparticles, satisfy the usual anticommutation relations. The
non-vanishing ones are 
\begin{equation}
\left\{ b_{\mathbf{p}s},b_{\mathbf{q}\sigma }^{\dagger }\right\} =\left\{ d_{%
\mathbf{p}s},d_{\mathbf{q}\sigma }^{\dagger }\right\} =\delta _{s\sigma
}\delta ^{3}\left( \mathbf{p}-\mathbf{q}\right) .
\end{equation}

Again, the analogous properties apply to the mass-$m_{2}$ operators $%
B^{\dagger }$, $B$, $D^{\dagger }$, $D$. As a concrete example, we can think
of the mass-$m_{1}$ particles as electrons, and the mass-$m_{2}$ particles
as muons, through any pairs of charged fermions could be considered.

\vskip0.8truecm {\normalsize \noindent {\textbf{\large 3. Variational
principle, two-fermion trial states and equations}}} {\normalsize \vskip%
0.4truecm }

The Hamiltonian formalism of QFT is based on the covariant eigenvalue
equation

\begin{equation}
\widehat{P}^{\mu }\,\left| \psi \right\rangle =Q^{\mu }\,\left| \psi
\right\rangle ,
\end{equation}
where $\widehat{P}^{\mu }=\left( \widehat{H},\widehat{\mathbf{P}}\right) $
is the energy momentum operator, $Q^{\mu }=\left( E,\mathbf{P}\right) $ is
the 4-vector of energy and momentum, with $E^{2}-\mathbf{P}^{2}\mathbf{=}%
M^{2}$, where $M$ is the invariant mass of the system. There are very few
problem in QFT for which exact solution of the $\mu =0$ equation (22) can be
written down. In practice, it is necessary to use approximation methods to
solve it, such as the widely used covariant perturbation theory or lattice
methods. Here we concentrate on the variational approach, which is based on
the variational principle 
\begin{equation}
\delta \,\left\langle \psi \right| \,\widehat{H}-E\,\left| \psi
\right\rangle _{t=0}=0,
\end{equation}
which we shall consider in the rest frame i.e. $\mathbf{P=}0$. Variational
solutions are only as good as the trial states that are used. Thus, it is
important that the trial states possess as many features of the exact
solution as possible. For a system like $\mu ^{+}e^{-}$, the simplest
Fock-space trial state that can be written down in the rest frame is

\begin{equation}
\left| \psi _{trial}\right\rangle =\underset{s_{1}s_{2}}{\sum }\int d^{3}%
\mathbf{p}F_{s_{1}s_{2}}(\mathbf{p})b_{\mathbf{p}s_{1}}^{\dagger }D_{-%
\mathbf{p}s_{2}}^{\dagger }\left| 0\right\rangle ,
\end{equation}
where $F_{s_{1}s_{2}}$ are four adjustable functions. We use this trial
state to evaluate the matrix elements needed to implement the variational
principle (23), that is 
\begin{equation}
\left\langle \psi _{trial}\right| :\widehat{H}_{0}-E:\left| \psi
_{trial}\right\rangle =\underset{s_{1}s_{2}}{\sum }\int d^{3}\mathbf{p}%
F_{s_{1}s_{2}}^{\ast }(\mathbf{p})F_{s_{1}s_{2}}(\mathbf{p})\left( \omega
_{p}+\Omega _{p}-E\right)
\end{equation}
and

\begin{equation*}
\left\langle \psi _{trial}\right| :\widehat{H}_{I}:\left| \psi
_{trial}\right\rangle =-\frac{q_{1}q_{2}m_{1}m_{2}}{\left( 2\pi \right) ^{3}}%
\underset{\sigma _{1}\sigma _{2}s_{1}s_{2}}{\sum }\int \frac{d^{3}\mathbf{p}%
d^{3}\mathbf{q}}{\sqrt{\omega _{p}\omega _{q}\Omega _{p}\Omega _{q}}}%
F_{s_{1}s_{2}}^{\ast }(\mathbf{p})F_{\sigma _{1}\sigma _{2}}(\mathbf{q}%
)\times
\end{equation*}
\begin{equation}
\times \overline{u}\left( \mathbf{p},s_{1}\right) \gamma ^{\mu }u\left( 
\mathbf{q},\sigma _{1}\right) D_{\mu \nu }(p-q)\overline{V}\left( -\mathbf{q}%
,\sigma _{2}\right) \gamma ^{\nu }V\left( -\mathbf{p,}s_{2}\right) ,
\end{equation}
where the Fourier transform of the Green function was used: 
\begin{equation}
D_{\mu \nu }(x-x^{\prime })=\int \frac{d^{4}k}{(2\pi )^{4}}D_{\mu \nu
}(k)e^{-ik(x-x^{\prime })}.
\end{equation}
For a particle-antiparticle system like positronium an additional
virtual-annihilation interaction term, 
\begin{equation}
\overline{u}\left( \mathbf{p},s_{1}\right) \gamma ^{\mu }v\left( -\mathbf{p,}%
s_{2}\right) D_{\mu \nu }\left( \omega _{p}\right) \overline{v}\left( -%
\mathbf{q},\sigma _{2}\right) \gamma ^{\nu }u\left( \mathbf{q},\sigma
_{1}\right)
\end{equation}
appears in (26) [10]. Note that\ we have normal-order the entire
Hamiltonian, since this circumvents the need for mass renormalization which
would otherwise arise. Not that there is a difficulty with handling mass
renormalization in the present formalism (as shown in various earlier
papers; see, for example, ref. [11] and citations therein). It is simply
that we are not interested in mass renormalization here, since it has no
effect on the two-body bound state energies that we obtain in this paper.
Furthermore, the approximate trial state (24), which we use in this work, is
incapable of sampling loop effects. Thus, the normal-ordering of the entire
Hamiltonian does not ``sweep under the carpet'' loop renormalization
effects, since none arise at the present level of approximation. Note, also,
that $\left\langle \psi _{trial}\right| :\widehat{H}_{II}:\left| \psi
_{trial}\right\rangle =0$, that is the variational trial state (24) is
insensitive to that part of the interaction Hamiltonian which is linear in $%
A_{0}^{\mu }\left( x\right) $. This means that, with the simple anzatz (24)
only stable bound states and elastic scattering can be described, but not
processes that involve radiation.

The variational principle (23) leads to the following equation 
\begin{eqnarray}
&&\sum_{s_{1}s_{2}}\int d^{3}\mathbf{p}\left( \omega _{p}+\Omega
_{p}-E\right) F_{s_{1}s_{2}}(\mathbf{p})\delta F_{s_{1}s_{2}}^{\ast }(%
\mathbf{p})  \notag \\
&&-\frac{m_{1}m_{2}}{\left( 2\pi \right) ^{3}}\underset{\sigma _{1}\sigma
_{2}s_{1}s_{2}}{\sum }\int \frac{d^{3}\mathbf{p}d^{3}\mathbf{q}}{\omega
_{p}\omega _{q}}F_{\sigma _{1}\sigma _{2}}(\mathbf{q})\left( -i\right) 
\mathcal{M}_{s_{1}s_{2}\sigma _{1}\sigma _{2}}^{ope}\left( \mathbf{p,q}%
\right) \delta F_{s_{1}s_{2}}^{\ast }(\mathbf{p})=0,
\end{eqnarray}
where $\mathcal{M}_{s_{1}s_{2}\sigma _{1}\sigma _{2}}^{ope}\left( \mathbf{p,q%
}\right) $ is the usual invariant matrix element corresponding to the
one-photon exchange Feynman diagram: 
\begin{equation}
\mathcal{M}_{s_{1}s_{2}\sigma _{1}\sigma _{2}}^{ope}\left( \mathbf{p,q}%
\right) =-\overline{u}\left( \mathbf{p},s_{1}\right) \left( -iq_{1}\gamma
^{\mu }\right) u\left( \mathbf{q},\sigma _{1}\right) iD_{\mu \nu }(p-q)%
\overline{V}\left( -\mathbf{q},\sigma _{2}\right) \left( -iq_{2}\gamma ^{\nu
}\right) V\left( -\mathbf{p},s_{2}\right) ,
\end{equation}
As mentioned above, for a fermion-antifermion system like positronium we
obtain [10] the additional virtual-annihilation term ($q_{1}=q_{2}\equiv e$) 
\begin{equation}
\mathcal{M}_{s_{1}s_{2}\sigma _{1}\sigma _{2}}^{ann}\left( \mathbf{p,q}%
\right) =\overline{u}\left( \mathbf{p},s_{1}\right) \left( -ie\gamma ^{\mu
}\right) v\left( -\mathbf{p},s_{2}\right) iD_{\mu \nu }\left( \omega
_{p}\right) \overline{v}\left( -\mathbf{q},\sigma _{2}\right) \left(
-ie\gamma ^{\nu }\right) u\left( \mathbf{q},\sigma _{1}\right) .
\end{equation}
Note that the $\mathcal{M}$-matrix arises naturally in this formalism, that
is $\mathcal{M}$ is not put in by hand, nor does its derivation require
additional Fock-space terms in the variational trial state (24), as is the
case in traditional formulations (e.g. [12], [13]).

In the non-relativistic limit, the functions $F_{s_{1}s_{2}}$ can be written
as 
\begin{equation}
F_{s_{1}s_{2}}(\mathbf{p})=F(\mathbf{p})\Lambda _{s_{1}s_{2}},
\end{equation}
where the non-zero elements of $\Lambda _{ij}$ for total spin singlet ($S=0$%
) states are $\Lambda _{12}=-\Lambda _{21}=\frac{1}{\sqrt{2}}$, while for
the spin triplet ($S=1$)\ states the non-zero elements are $\Lambda _{11}=1$
for $m_{s}=+1,$ $\Lambda _{12}=\Lambda _{21}=\frac{1}{\sqrt{2}}$ for $%
m_{s}=0 $, and $\Lambda _{22}=1$ for $m_{s}=-1$. We use the notation that
the subscripts 1 and 2 of $\Lambda $ correspond to $m_{s}=1/2$ and $%
m_{s}=-1/2$ (or $\uparrow $\ and\ $\downarrow $) respectively. Substituting
(32) into (29), multiplying the result by $\Lambda _{s_{1}s_{2}}$ and
summing over $s_{1}$ and $s_{2}$, gives the equation 
\begin{equation}
(\omega (\mathbf{p})+\Omega (\mathbf{p})-E)F(\mathbf{p})=\frac{1}{(2\pi )^{3}%
}\int d^{3}\mathbf{q}\mathcal{K}(\mathbf{p},\mathbf{q})F(\mathbf{q}),
\end{equation}
where 
\begin{equation}
\mathcal{K}(\mathbf{p},\mathbf{q})=-i\sum_{s_{1}s_{2}\sigma _{1}\sigma
_{2}}\Lambda _{s_{1}s_{2}}\mathcal{M}_{s_{1}s_{2}\sigma _{1}\sigma
_{2}}\left( \mathbf{p,q}\right) \Lambda _{\sigma _{1}\sigma _{2}}.
\end{equation}
To lowest-order in $\left( \left| \mathbf{p}\right| \mathbf{,}\left| \mathbf{%
q}\right| \right) /\left( m_{1},m_{2}\right) $ (i.e. in the non-relativistic
limit), the kernel (34) reduces to $\mathcal{K}=q_{1}q_{2}/\left| \mathbf{p-q%
}\right| ^{2}$, and so (33) reduces to the (momentum-space) Schr\"{o}dinger
equation 
\begin{equation}
\left( \frac{\mathbf{p}^{2}}{2m_{r}}-\varepsilon \right) F(\mathbf{p})=\frac{%
q_{1}q_{2}}{(2\pi )^{3}}\int d^{3}\mathbf{q}\frac{1}{\left| \mathbf{p-q}%
\right| ^{2}}F(\mathbf{q}),
\end{equation}
where $\varepsilon =E-M$ and $m_{r}=m_{1}m_{2}/M$, $M=m_{1}+m_{2}$. This
verifies that the relativistic two-fermion equation (29) has the required
non-relativistic limit.

\vskip0.8truecm {\normalsize \noindent {\large 4}{\textbf{\large .
Partial-wave decomposition and classification of states}}} {\normalsize %
\vskip0.4truecm }

In the relativistic case we shall not complete the variational procedure in
(29) at this stage to get final equations for the four functions $%
F_{s_{1}s_{2}}$, because they are not independent in general. We require
that the trial state must be an eigenstate of the relativistic total angular
momentum operator, its projection, and parity. Namely 
\begin{equation}
\left[ 
\begin{array}{c}
\widehat{\mathbf{J}}^{2} \\ 
\widehat{J}_{3} \\ 
\widehat{\mathcal{P}}
\end{array}
\right] \,\left| \psi _{trial}\right\rangle =\left[ 
\begin{array}{c}
J\left( J+1\right) \\ 
m_{J} \\ 
P
\end{array}
\right] \,\left| \psi _{trial}\right\rangle .
\end{equation}
For a system like positronium charge conjugation invariance is an additional
requirement, that is 
\begin{equation}
\widehat{\mathcal{C}}\,\left| e^{+}e^{-}\right\rangle =C\,\left|
e^{+}e^{-}\right\rangle ,
\end{equation}
however this does not apply for $m_{1}\neq m_{2}$.\ Explicit forms for the
operators $\widehat{\mathbf{J}}^{2}$, $\widehat{J}_{3}$ are given in
Appendix A. The functions $F_{s_{1}s_{2}}(\mathbf{p})$ can be written in the
general form 
\begin{equation}
F_{s_{1}s_{2}}(\mathbf{p})=\sum_{\ell
_{s_{1}s_{2}}}\sum_{m_{s_{1}s_{2}}}f_{s_{1}s_{2}}^{\ell
_{s_{1}s_{2}}m_{s_{1}s_{2}}}\left( p\right) Y_{\ell
_{s_{1}s_{2}}}^{m_{s_{1}s_{2}}}(\widehat{\mathbf{p}}),
\end{equation}
where $Y_{\ell _{s_{1}s_{2}}}^{m_{s_{1}s_{2}}}(\widehat{\mathbf{p}})$\ are
the usual spherical harmonics. Here and henceforth we will use the notation $%
p=\left| \mathbf{p}\right| $ etc. (four-vectors will be written as $p^{\mu }$%
). The orbital indices $\ell _{s_{1}s_{2}}$and $m_{s_{1}s_{2}}$ depend on
the spin indices $s_{1}$ and $s_{2}$\ and are specified by equations (36).
The radial coefficients of expansion (38) $f_{s_{1}s_{2}}^{\ell
_{s_{1}s_{2}}m_{s_{1}s_{2}}}\left( p\right) $ also depend on the spin
variables. Substitution of (38) into (24) and then into (36) leads to two
categories of relations among the adjustable functions.

\vskip0.2truecm

\textbf{Mixed-spin states }

In this case $\ell _{s_{1}s_{2}}\equiv \ell =J$ and the general solution of
the system (36) is 
\begin{equation}
F_{s_{1}s_{2}}(\mathbf{p})=C_{1}F_{s_{1}s_{2}}^{\left( sg\right) }(\mathbf{p}%
)+C_{2}F_{s_{1}s_{2}}^{\left( tr\right) }(\mathbf{p}),
\end{equation}
where $C_{1}$ and $C_{2}$ are arbitrary constants. $F_{s_{1}s_{2}}^{\left(
sg\right) }(\mathbf{p})$ and $F_{s_{1}s_{2}}^{\left( tr\right) }(\mathbf{p})$
are functions, which correspond to pure singlet states with the total spin $%
S=0$ and triplet states with $S=1$ respectively. The singlet functions have
the form 
\begin{equation}
F_{s_{1}s_{2}}^{\left( sg\right) }(\mathbf{p})=f_{s_{1}s_{2}}^{\left(
sg\right) \ell }(p)Y_{\ell }^{m_{s_{1}s_{2}}}(\widehat{\mathbf{p}}),
\end{equation}
where $m_{11}=m_{22}=0\ $and$\;m_{12}=m_{21}=m_{J}$.\ The relations between
\ $f_{12}^{\left( sg\right) J}(p)\;$and$\;\,f_{21}^{\left( sg\right) J}(p)$
involve the Clebsch-Gordon (C-G) coefficients $C^{\left( sg\right)
m_{s_{1}s_{2}}}$, that is 
\begin{equation}
f_{s_{1}s_{2}}^{\left( sg\right) J}(p)=C^{\left( sg\right)
m_{s_{1}s_{2}}}f^{J}(p),
\end{equation}
as it shown in Appendix A. We see that the spin and radial variables
separate in the sense that the factors $f_{s_{1}s_{2}}^{(J)}(p)$ have a
common radial function $f^{J}(p),$ that is for the singlet functions we
obtain 
\begin{equation}
F_{s_{1}s_{2}}^{\left( sg\right) }(\mathbf{p})=C^{\left( sg\right)
m_{s_{1}s_{2}}}f^{J}(p)Y_{J}^{m_{s_{1}s_{2}}}(\widehat{\mathbf{p}}).
\end{equation}
The C-G coefficients $C^{\left( sg\right) m_{s_{1}s_{2}}}$\ have a simple
form: $C^{\left( sg\right) m_{11}}=C^{\left( sg\right) m_{22}}=0$,\ $%
C^{\left( sg\right) m_{12}}=-C^{\left( sg\right) m_{21}}=1$ (see Appendix
A). Thus the nonzero components of $F_{s_{1}s_{2}}^{\left( sg\right) }(%
\mathbf{p})$ are $F_{\uparrow \downarrow }^{\left( sg\right) }(\mathbf{p}%
)\equiv F_{12}^{\left( sg\right) }(\mathbf{p}),\;F_{\downarrow \uparrow
}^{\left( sg\right) }(\mathbf{p})\equiv F_{21}^{\left( sg\right) }(\mathbf{p}%
)$.

The triplet functions have the form

\begin{equation}
F_{s_{1}s_{2}}^{\left( tr\right) }(\mathbf{p})=f_{s_{1}s_{2}}^{\left(
tr\right) J}(p)Y_{J}^{m_{s_{1}s_{2}}}(\widehat{\mathbf{p}}),
\end{equation}
where 
\begin{equation}
m_{11}=m_{J}-1,\;\ \ \ m_{12}=m_{21}=m_{J},\;\ \ \ m_{22}=m_{J}+1.
\end{equation}
The expressions for $f_{s_{1}s_{2}}^{J}(p)$ involve the C-G coefficients $%
C_{Jm_{J}}^{\left( tr\right) Jm_{s}}$ for $S=1$ listed in Appendix A, that
is 
\begin{equation}
f_{s_{1}s_{2}}^{\left( tr\right) J}(p)=C_{Jm_{J}}^{\left( tr\right)
Jm_{s}}f^{\ell }(p),
\end{equation}
where the index $m_{s}$ is defined as 
\begin{eqnarray}
m_{s} &=&+1,\;\ \ when\;\ \ m_{s_{1}s_{2}}=m_{11},  \notag \\
m_{s} &=&0,\;\ \ \ \ when\;\ \ \ m_{s_{1}s_{2}}=m_{12}=m_{21}, \\
m_{s} &=&-1,\;\ \ when\;\ \ m_{s_{1}s_{2}}=m_{22}.  \notag
\end{eqnarray}
Thereupon, the triplet function is 
\begin{equation}
F_{s_{1}s_{2}}^{\left( tr\right) }(\mathbf{p})=C_{Jm_{J}}^{\left( tr\right)
Jm_{s}}f^{J}(p)Y_{J}^{m_{s_{1}s_{2}}}(\widehat{\mathbf{p}}).
\end{equation}
We need to note that (42) is true for $J\geq 0$, while (47) is true for $%
J\geq 1$. Thus the coefficient $C_{2}$ in (39) is zero when $J=0$. In other
words, for $J=0$, only the pure singlet state arises. For a system like
positronium the requirement (37) decouples the singlet and triplet states
for all $J$. Indeed, the charge conjugation eigenstates are 
\begin{equation}
\left| sg\right\rangle =\sum_{_{s_{1}s_{2}}}C^{\left( sg\right)
m_{s_{1}s_{2}}}\int d^{3}\mathbf{p\,}f^{J}(p)Y_{J}^{m_{J}}(\widehat{\mathbf{p%
}})b_{\mathbf{p}s_{1}}^{\dagger }D_{-\mathbf{p}s_{2}}^{\dagger }\mid 0\rangle
\end{equation}
with $C=\left( -1\right) ^{J}$ for the pure singlet states, and 
\begin{equation}
\left| tr\right\rangle =\sum_{_{s_{1}s_{2}}}C_{Jm_{J}}^{\left( tr\right)
Jm_{s_{1}s_{2}}}\int d^{3}\mathbf{p\,}f^{J}(p)Y_{J}^{m_{J}}(\widehat{\mathbf{%
p}})b_{\mathbf{p}s_{1}}^{\dagger }D_{-\mathbf{p}s_{2}}^{\dagger }\mid
0\rangle
\end{equation}
with $C=\left( -1\right) ^{J+1}$ for the pure triplet states, as it
discussed in Appendix A.

The states (48) and (49) diagonalize the Hamiltonian (13) only for
particle-antiparticle systems when $m_{1}=m_{2}$. Thus, for positronium-like
systems, the states can be characterized by the spin quantum number $S$, and
the mixed states (39) separate into singlet states (parastates $S=0$) and
triplet states (orthostates $S=1$). For distinct particles ($m_{1}\neq m_{2}$%
) $C$ is not conserved and there is no separation into pure singlet and
triplet states in general. Thus for arbitrary mass ratios we need to
diagonalize the expectation value of the Hamiltonian (13) (Appendix E). This
can be achieved by the following linear transformation\ 

\begin{equation}
\left[ 
\begin{array}{c}
\left| sg_{q}\right\rangle \\ 
\left| tr_{q}\right\rangle
\end{array}
\right] =\widehat{U}\left[ 
\begin{array}{c}
\left| sg\right\rangle \\ 
\left| tr\right\rangle
\end{array}
\right] ,
\end{equation}
where $\widehat{U}$ is the unimodular matrix 
\begin{equation}
\widehat{U}=\left[ 
\begin{array}{cc}
a & -b \\ 
b & a
\end{array}
\right]
\end{equation}
with components 
\begin{equation}
a=\sqrt{\frac{1+\xi }{2}},\;\;\;\;\;\;\;\;\;\;b=\sqrt{\frac{1-\xi }{2}},
\end{equation}
where 
\begin{equation}
\xi =\left( 4\left( \left( m_{1}-m_{2}\right) /\left( m_{1}+m_{2}\right)
\right) ^{2}J\left( J+1\right) +1\right) ^{-1/2}.
\end{equation}
The new states, which diagonalize the expectation value of $\widehat{H}$,
shall be called quasi-singlet $\mid sg_{q}\rangle $ and quasi-triplet $\mid
tr_{q}\rangle $ states 
\begin{eqnarray}
\left| sg_{q}\right\rangle &=&\sum_{_{s_{1}s_{2}}}C_{Jm_{J}}^{\left(
s_{s}\right) Jm_{s_{1}s_{2}}}\int d^{3}\mathbf{p\,}f^{J}(p)Y_{J}^{m_{J}}(%
\widehat{\mathbf{p}})b_{\mathbf{p}s_{1}}^{\dagger }D_{-\mathbf{p}%
s_{2}}^{\dagger }\mid 0\rangle ,  \notag \\
&& \\
\left| tr_{q}\right\rangle &=&\sum_{_{s_{1}s_{2}}}C_{Jm_{J}}^{\left(
s_{t}\right) Jm_{s_{1}s_{2}}}\int d^{3}\mathbf{p\,}f^{J}(p)Y_{J}^{m_{J}}(%
\widehat{\mathbf{p}})b_{\mathbf{p}s_{1}}^{\dagger }D_{-\mathbf{p}%
s_{2}}^{\dagger }\mid 0\rangle ,  \notag
\end{eqnarray}
where the coefficients $C_{Jm_{J}}^{\left( s_{s}\right) Jm_{s_{1}s_{2}}}$
and $C_{Jm_{J}}^{\left( s_{t}\right) Jm_{s_{1}s_{2}}}$ are listed in Table
1, and satisfy the following condition 
\begin{equation}
\sum_{\nu _{1}\nu _{2}m_{J}}\left( C_{Jm_{J}}^{\left( s_{s}\right) Jm_{\nu
}}\right) ^{2}=\sum_{\nu _{1}\nu _{2}m_{J}}\left( C_{Jm_{J}}^{\left(
s_{t}\right) Jm_{\nu }}\right) ^{2}=2\left( 2J+1\right) .
\end{equation}

{\normalsize \vskip0.4truecm }

\begin{center}
Table 1. The quasi-state coefficients $C_{Jm_{J}}^{\left( s\right)
Jm_{s_{1}s_{2}}}$ for $s=s_{s},\;s_{t}$.

\begin{tabular}{||c|c|c|c|c||}
\hline
& $m_{11}=m_{J}-1$ & $m_{12}=m_{J}$ & $m_{21}=m_{J}$ & $m_{22}=m_{J}+1$ \\ 
\hline
$s_{s}$ & $b\left( \frac{\left( J+m_{J}\right) \left( J-m_{J}+1\right) }{%
J\left( J+1\right) }\right) ^{1/2}$ & $-b\frac{m_{J}}{\left( J\left(
J+1\right) \right) ^{1/2}}+a$ & $-b\frac{m_{J}}{\left( J\left( J+1\right)
\right) ^{1/2}}-a$ & $-b\left( \frac{\left( J-m_{J}\right) \left(
J+m_{J}+1\right) }{J\left( J+1\right) }\right) ^{1/2}$ \\ \hline
$s_{t}$ & $-a\left( \frac{\left( J+m_{J}\right) \left( J-m_{J}+1\right) }{%
J\left( J+1\right) }\right) ^{1/2}$ & $a\frac{m_{J}}{\left( J\left(
J+1\right) \right) ^{1/2}}+b$ & $a\frac{m_{J}}{\left( J\left( J+1\right)
\right) ^{1/2}}-b$ & $a\left( \frac{\left( J-m_{J}\right) \left(
J+m_{J}+1\right) }{J\left( J+1\right) }\right) ^{1/2}$ \\ \hline
\end{tabular}
\end{center}

Note that these coefficients differ from C-G coefficients, because of the
coupled system we are dealing with. Remember that these coupled quasi-states
arise only for $J>0$. For $J=0$ purely $S=0$ states occur. Quasi-singlet and
quasi-triplet states are both characterized by the same quantum numbers $J$, 
$m_{J}$ and $P=(-1)^{J+1}$. Because of the unimodularity of matrix (51) we
can identify quasi-singlet and quasi-triplet states by quasi-spin (like
isospin) $t=1/2$ with $t_{3}=\mp 1/2$, which is a new quantum number (or
label). However, for our purpose it is more convenient to use the value $%
s=t_{3}+1/2$, which gives $s=s_{s}=0$ or $s=s_{t}=1$ for quasi-singlet and
quasi-triplet states respectively. In this case the labels $s_{s}$ and $%
s_{t} $ reflect better the meaning of the indicated\ quasi-states. It is
easy to see from (54) and Table 1 that for positronium the quasi states
become true singlet ($b=0$) and triplet ($a=0$) states with different charge
conjugation quantum numbers. It is useful to note for subsequent
calculations that the coefficients $C_{1}$ and $C_{2}$ in (39) are $C_{1}=a$%
, $C_{2}=-b$ for quasi-singlet states $s=s_{s}=0$, and $C_{1}=b$, $C_{2}=a$
for quasi-triplet states $s=s_{t}=1$.

\vskip0.2truecm

\textbf{The triplet }$\ell $-\textbf{mixing states}

These states occur for $\ell _{s_{1}s_{2}}\equiv \ell =J\mp 1$ (see Appendix
A). The adjustable functions have the form 
\begin{equation}
F_{s_{1}s_{2}}(\mathbf{p})=C_{Jm_{J}}^{\left( tr\right)
(J-1)m_{s}}f^{J-1}(p)Y_{J-1}^{m_{s_{1}s_{2}}}(\widehat{\mathbf{p}}%
)+C_{Jm_{J}}^{\left( tr\right) (J+1)m_{s}}f^{J+1}(p)Y_{J+1}^{m_{s_{1}s_{2}}}(%
\widehat{\mathbf{p}}),
\end{equation}
where $m_{s_{1}s_{2}}$ are defined in (46), while the coefficients $%
C_{Jm_{J}}^{\left( tr\right) (J\mp 1)m_{s}}$, which are precisely the C-G
coefficients, can be found in Appendix A. Expression (56) involves two
radial functions $f^{J-1}(p)$ and $f^{J+1}(p)$ which correspond to $\ell
=J-1 $ and $\ell =J+1$. This reflects the fact that the orbital angular
momentum is not conserved and $\ell $ is not a good quantum number. The
system in these states is characterized by $J,$ $m_{J},$ and $P=(-1)^{J}$.
In spectroscopic notation, these states are a mixture of $^{3}\left(
J-1\right) _{J}$, and $^{3}\left( J+1\right) _{J}$\ states. The exception is
the state with $J=0$, for which the orbital angular momentum is a good
quantum number. Indeed, for $J=0$\ the function $f^{J-1}(p)$ does not exist
(see Appendix A), thus the function $F_{s_{1}s_{2}}(\mathbf{p})$ is defined
only by the second term in (56). Note that $\ell $-mixing states appear for
principal quantum number $n\geq 3$\ only.

\vskip0.8truecm {\normalsize \noindent {\large 5}{\textbf{\large . The
relativistic radial equations for two fermion systems}}} {\normalsize \vskip%
0.4truecm }

It is not possible to write an universal two fermion wave equation, because
the adjustable functions have different form for different states. Thus it
was important to classify all states of the system before deriving final
radial equations. Now we return to the variational equation (29) and replace
the functions $F_{s_{1}s_{2}}(\mathbf{p})$ by the expression (39) for the
quasi-states and by (56) for the triplet states. After completing the
variational procedure we obtain the following results:

For the states with $\ell =J$, $P=(-1)^{J+1}$ the radial equations are

\begin{equation}
\left( \omega _{p}+\Omega _{p}-E\right) f^{J}(p)=\frac{m_{1}m_{2}}{\left(
2\pi \right) ^{3}}\int \frac{q^{2}dq}{\sqrt{\omega _{p}\omega _{q}\Omega
_{p}\Omega _{q}}}\mathcal{K}\left( p,q\right) f^{J}(q),
\end{equation}
where the kernel $\mathcal{K}\left( p,q\right) $\ is defined by invariant $%
\mathcal{M}$-matrices as follows from (29). For the pure singlet states with 
$J=0$, according to (42), (48) and (147), the kernel $\mathcal{K}\left(
p,q\right) $\ is

\begin{equation}
\mathcal{K}\left( p,q\right) =-\frac{i}{4\pi }\int d\widehat{\mathbf{p}}\,d%
\widehat{\mathbf{q}}\left( \mathcal{M}_{1212}\left( \mathbf{p,q}\right) -%
\mathcal{M}_{1221}\left( \mathbf{p,q}\right) -\mathcal{M}_{2112}\left( 
\mathbf{p,q}\right) +\mathcal{M}_{2121}\left( \mathbf{p,q}\right) \right) ,
\end{equation}
For quasi-singlet and quasi-triplet states $\left( J\geq 1\right) $ we have

\begin{equation}
\mathcal{K}\left( p,q\right) =\underset{s_{1}s_{2}\sigma _{1}\sigma _{2}m_{J}%
}{-i\sum }\int d\widehat{\mathbf{p}}\,d\widehat{\mathbf{q}}%
\,C_{Jm_{J}}^{\left( s\right) s_{1}s_{2}\sigma _{1}\sigma _{2}}\mathcal{M}%
_{s_{1}s_{2}\sigma _{1}\sigma _{2}}\left( \mathbf{p,q}\right)
Y_{J}^{m_{J}\ast }(\widehat{\mathbf{p}})Y_{J}^{m_{J}}(\widehat{\mathbf{q}}),
\end{equation}
where the coefficients $C_{Jm_{J}}^{\left( s\right) s_{1}s_{2}\sigma
_{1}\sigma _{2}}$ are expressed through the coefficients $C_{Jm_{J}}^{\left(
s\right) Jm_{s_{1}s_{2}}}$ 
\begin{equation}
C_{Jm_{J}}^{\left( s\right) s_{1}s_{2}\sigma _{1}\sigma _{2}}\equiv
C_{Jm_{J}}^{\left( s\right) Jm_{s_{1}s_{2}}}C_{Jm_{J}}^{\left( s\right)
Jm_{\sigma _{1}\sigma _{2}}}/\sum_{\nu _{1}\nu _{2}m_{J}}\left(
C_{Jm_{J}}^{\left( s\right) Jm_{\nu _{1}\nu _{2}}}\right) ^{2},
\end{equation}
where $s=s_{s},s_{t}$\ .\ In (60) we have summed over $m_{J}$, because of
the $2J+1$ energy degeneracy.

For the triplet states with $\ell =J\mp 1$, we have two independent radial
functions $f^{J-1}(p)\;$and$\;f^{J+1}(p)$.\ Thus the variational equation
(29) leads to a system of coupled equations for $f^{J-1}(p)\;$and$%
\;f^{J+1}(p)$. It is convenient to write them in matrix form,

\begin{equation}
\left( \omega _{p}+\Omega _{p}-E\right) \mathbb{F}\left( p\right) =\frac{%
m_{1}m_{2}}{\left( 2\pi \right) ^{3}}\int \frac{q^{2}dq}{\sqrt{\omega
_{p}\omega _{q}\Omega _{p}\Omega _{q}}}\mathbb{K}\left( p,q\right) \mathbb{F}%
\left( q\right) ,
\end{equation}
where 
\begin{equation}
\mathbb{F}\left( p\right) =\left[ 
\begin{array}{c}
f^{J-1}(p) \\ 
f^{J+1}(p)
\end{array}
\right] ,
\end{equation}
and 
\begin{equation}
\mathbb{K}\left( p,q\right) =\left[ 
\begin{array}{cc}
\mathcal{K}_{11}\left( p,q\right) & \mathcal{K}_{12}\left( p,q\right) \\ 
\mathcal{K}_{21}\left( p,q\right) & \mathcal{K}_{22}\left( p,q\right)
\end{array}
\right] .
\end{equation}
The kernels $\mathcal{K}_{ij}$ are similar in form to (59), that is 
\begin{equation}
\mathcal{K}_{ij}\left( p,q\right) =\underset{s_{1}s_{2}\sigma _{1}\sigma
_{2}m_{J}}{-i\sum }C_{Jm_{J}ij}^{s_{1}s_{2}\sigma _{1}\sigma _{2}}\int d%
\widehat{\mathbf{p}}\,d\widehat{\mathbf{q}}\,\mathcal{M}_{s_{1}s_{2}\sigma
_{1}\sigma _{2}}\left( \mathbf{p,q}\right) Y_{\ell _{i}}^{m_{s_{1}s_{2}}\ast
}(\widehat{\mathbf{p}})Y_{\ell _{j}}^{m_{\sigma _{1}\sigma _{2}}}(\widehat{%
\mathbf{q}}).
\end{equation}
However the coefficients $C_{Jm_{J}ij}^{s_{1}s_{2}\sigma _{1}\sigma _{2}}$
are defined by the expression 
\begin{equation}
C_{Jm_{J}ij}^{s_{1}s_{2}\sigma _{1}\sigma _{2}}=C_{Jm_{J}}^{\left( tr\right)
\ell _{i}m_{s_{1}s_{2}}}C_{Jm_{J}}^{\left( tr\right) \ell _{j}m_{\sigma
_{1}\sigma _{2}}}/\sum_{s_{1}s_{2}m_{J}}\left( C_{Jm_{J}}^{\left( tr\right)
\ell _{i}m_{s}}\right) ^{2},
\end{equation}
where $\ell _{1}=J-1,\;\ell _{2}=J+1.$ The system (61) reduces to a single
equation for $f^{J+1}(p)$ when $J=0$, since $f^{J-1}(p)=0$ in that case.

To our knowledge, it is not possible to obtain analytic solution of the
relativistic radial momentum-space equations (57) and (61). Thus one must
resort to numerical or other approximation methods. Numerical solutions of
such equations are discussed, for example, in [11], while a variational
approximations have been employed in [5]. However, in this paper we shell
resort to perturbative approximations, in order to verify that our\
equations agree with known results for positronium to $O(\alpha ^{4})$.\ We
expect that this must be so, given that the interaction kernels (i.e.
momentum-space potentials) of our equations involve the ``tree-level''
Feynman diagrams only.

Beyond $O(\alpha ^{4})$ our equations are evidently incomplete. One could,
of course, augment them by the addition of invariant matrix elements
corresponding to one-loop Feynman diagrams to the existing $\mathcal{M}$%
-matrices in the kernels of our equations. Indeed such an approach has been
used in a similar though not variational treatment of positronium and
muonium by Zhang and Koniuk [14, 15]. These authors show that the inclusion
of single-loop diagrams yields positronium energy eigenstates which are
accurate to $O\left( \alpha ^{5},\alpha ^{5}\ln \alpha \right) $. However
such ad-hoc augmentation of the kernels would be contrary to the spirit of
the present variational treatment, and we shall not pursue it in this work.

\vskip0.8truecm {\normalsize \noindent {\large 6}{\textbf{\large . The
kernels in semi-relativistic expansion and the non-relativistic limit}} } 
{\normalsize \vskip0.4truecm }

For perturbative solutions of our radial equations, it is necessary to work
out expansions of the relevant expressions to first order beyond the
non-relativistic limit. This shall be summarized in the present section. We
shall do the calculation in the Coulomb gauge, in which the photon
propagator has the form [16] 
\begin{equation}
D_{00}\left( \mathbf{k}\right) =\frac{1}{\mathbf{k}^{2}},\;\;D_{0l}\left( 
\mathbf{k}\right) =0,\;\;D_{ij}\left( k^{\mu }\right) =\frac{1}{k^{\mu
}k_{\mu }}\left( \delta _{ij}-\frac{k_{i}k_{j}}{\mathbf{k}^{2}}\right) ,
\end{equation}
where $k^{\mu }=\left( \omega _{p}-\omega _{q},\mathbf{p-q}\right) $.

To expand the amplitudes $\mathcal{M}$\ of (30), (31)\ up to the lowest
non-trivial order of $\left( p/m\right) ^{2}$, we take the free-particle
spinors to be 
\begin{equation}
u(\mathbf{p,}i)=\left[ 
\begin{array}{c}
\left( 1+\frac{\mathbf{p}^{2}}{8m^{2}}\right) \\ 
\frac{(\overrightarrow{\sigma }\cdot \mathbf{p})}{2m}
\end{array}
\right] \varphi _{i},\;\ \ \ \ \ \ \ v(\mathbf{p,}i)=\left[ 
\begin{array}{c}
\frac{(\overrightarrow{\sigma }\mathbf{\cdot p})}{2m} \\ 
\left( 1+\frac{\mathbf{p}^{2}}{8m^{2}}\right)
\end{array}
\right] \chi _{i},
\end{equation}
as discussed in Appendix C. Analogously, the photon propagator takes on the
form

\begin{equation}
D_{00}\left( \mathbf{p-q}\right) =\frac{1}{\left( \mathbf{p-q}\right) ^{2}}%
,\;\;\;D_{kl}\left( \mathbf{p-q}\right) \simeq -\frac{1}{\left( \mathbf{p-q}%
\right) ^{2}}\left( \delta _{kl}-\frac{\left( p-q\right) _{k}\left(
p-q\right) _{l}}{\left( \mathbf{p-q}\right) ^{2}}\right) .
\end{equation}
Corresponding calculations give for the orbital part of the $\mathcal{M}$%
-matrix

\begin{eqnarray}
&&\mathcal{M}_{s_{1}s_{2}\sigma _{1}\sigma _{2}}^{ope\left( orb\right) }(%
\mathbf{q},\mathbf{p})  \notag \\
&=&iq_{1}q_{2}\left\{ \frac{1}{\left( \mathbf{p-q}\right) ^{2}}+\frac{1}{%
m_{1}m_{2}}\left( \frac{1}{2}\left( \frac{m_{1}}{m_{2}}+\frac{m_{2}}{m_{1}}%
\right) \left( \frac{1}{4}+\frac{\mathbf{q\cdot p}}{\left( \mathbf{p-q}%
\right) ^{2}}\right) +\frac{\left( \mathbf{p\times q}\right) ^{2}}{\left( 
\mathbf{p-q}\right) ^{4}}\right) \right\} \delta _{s_{1}\sigma _{1}}\delta
_{s_{2}\sigma _{2}}.
\end{eqnarray}
The linear spin terms, which are responsible for spin-orbit interaction,
become

\begin{equation}
\mathcal{M}_{s_{1}s_{2}\sigma _{1}\sigma _{2}}^{ope\left( s-o\right) }(%
\mathbf{q},\mathbf{p})=-\frac{q_{1}q_{2}}{4m_{1}m_{2}}\varphi
_{s_{1}}^{\dagger }\chi _{\sigma _{2}}^{\dagger }\left( \left( \frac{m_{2}}{%
m_{1}}+2\right) \overrightarrow{\sigma }_{1}-\left( \frac{m_{1}}{m_{2}}%
+2\right) \overrightarrow{\sigma }_{2}\right) \cdot \frac{\mathbf{p\times q}%
}{\left( \mathbf{p-q}\right) ^{2}}\varphi _{\sigma _{1}}\chi _{s_{2}}.
\end{equation}
Here $\overrightarrow{\sigma }_{1}$ and $\overrightarrow{\sigma }_{2}$\ are
spin matrices of first and second particles respectively, which are defined
by the following operations $\overrightarrow{\sigma }_{1}\varphi _{\sigma
_{1}}\chi _{s_{2}}=\left( \overrightarrow{\sigma }_{1}\varphi _{\sigma
_{1}}\right) \chi _{s_{2}}$, $\overrightarrow{\sigma }_{2}\varphi _{\sigma
_{1}}\chi _{s_{2}}=\varphi _{\sigma _{1}}\left( \overrightarrow{\sigma }%
_{2}\chi _{s_{2}}\right) $. The quadratic spin terms or spin-spin
interaction terms are

\begin{equation}
\mathcal{M}_{s_{1}s_{2}\sigma _{1}\sigma _{2}}^{ope\left( s-s\right) }(%
\mathbf{q},\mathbf{p})=\frac{iq_{1}q_{2}}{4m_{1}m_{2}}\varphi
_{s_{1}}^{\dagger }\chi _{\sigma _{2}}^{\dagger }\left\{ -\frac{\left( 
\overrightarrow{\sigma }_{2}\cdot \left( \mathbf{p-q}\right) \right) \left( 
\overrightarrow{\sigma }_{1}\cdot \left( \mathbf{p-q}\right) \right) }{%
\left( \mathbf{p-q}\right) ^{2}}+\left( \overrightarrow{\sigma }_{1}\cdot 
\overrightarrow{\sigma }_{2}\right) \right\} \varphi _{\sigma _{1}}\chi
_{s_{2}}.
\end{equation}
The annihilation contribution, which arises for a particle-antiparticle
system [10], is given by the term

\begin{equation}
\mathcal{M}_{s_{1}s_{2}\sigma _{1}\sigma _{2}}^{ann}(\mathbf{p},\mathbf{q})=-%
\frac{ie^{2}}{4m^{2}}\varphi _{s_{1}}^{\dagger }\chi _{\sigma _{2}}^{\dagger
}\left\{ \overrightarrow{\sigma }_{1}\cdot \overrightarrow{\sigma }%
_{2}\right\} \varphi _{\sigma _{1}}\chi _{s_{2}},
\end{equation}
where we have excluded a divergent term, which appears in the Coulomb gauge.
Here $m_{1}=m_{2}\equiv m$ and $q_{1}=q_{2}\equiv e$.

The kernels are calculated from (58), (59), (64) with (69)-(71). They
consist of three parts, namely

\begin{equation}
\mathcal{K}\left( p,q\right) =\mathcal{K}^{\left( orb\right) }\left(
p,q\right) +\mathcal{K}^{\left( s-o\right) }\left( p,q\right) +\mathcal{K}%
^{\left( s-s\right) }\left( p,q\right) .
\end{equation}

\vskip0.2truecm

\textbf{The singlet state} $\left( \ell =J=0,\;P=-1,\right) $

Details of the calculations can be found in Appendix D. We use the\
notation: $z=\left( p^{2}+q^{2}\right) /2pq$, and\ $Q_{\lambda }(z)$ are the
Legendre functions of the second kind [17]. The contributions of the various
terms to the kernel are as follows:

Orbital term 
\begin{eqnarray}
\mathcal{K}^{\left( sgl\right) \left( orb\right) }\left( p,q\right) &=&\frac{%
2\pi q_{1}q_{2}}{pq}Q_{0}(z)  \notag \\
&&+\frac{\pi q_{1}q_{2}}{2m_{1}m_{2}}\left( \left( \frac{m_{1}}{m_{2}}+\frac{%
m_{2}}{m_{1}}+1\right) \left( \frac{p}{q}+\frac{q}{p}\right)
Q_{0}(z)+2Q_{1}(z)-\left( \frac{m_{1}}{m_{2}}+\frac{m_{2}}{m_{1}}+2\right)
\right) ,
\end{eqnarray}

Spin-orbit interaction

\begin{equation}
\mathcal{K}^{\left( sgl\right) \left( s-o\right) }\left( p,q\right) =0,
\end{equation}

Spin-spin interaction

\begin{equation}
\mathcal{K}^{\left( sgl\right) \left( s-s\right) }\left( p,q\right) =\frac{%
2\pi q_{1}q_{2}}{m_{1}m_{2}}.
\end{equation}

\vskip0.2truecm

\textbf{The quasi-states} $\left( \ell =J\ (J\geq
1),\;P=(-1)^{J+1},\;s=0,1\right) $

The contributions of the various terms to the kernel are as follows:

Orbital term 
\begin{eqnarray}
\mathcal{K}^{\left( orb\right) }\left( p,q\right) &=&\frac{\pi q_{1}q_{2}}{pq%
}Q_{J}(z)  \notag \\
&&+\frac{\pi q_{1}q_{2}}{2m_{1}m_{2}}\left( \left( \frac{m_{1}}{m_{2}}+\frac{%
m_{2}}{m_{1}}-\left( J-1\right) \right) \left( \frac{p}{q}+\frac{q}{p}%
\right) Q_{J}(z)+2\left( J+1\right) Q_{J+1}(z)\right) ,
\end{eqnarray}

Spin-orbit interaction

\begin{eqnarray}
\mathcal{K}^{\left( s-o\right) }(p,q) &=&\frac{\pi q_{1}q_{2}}{2m_{1}m_{2}}%
\left( -C_{2}^{2}\left( \frac{m_{1}}{m_{2}}+\frac{m_{2}}{m_{1}}+4\right)
+2C_{1}C_{2}\sqrt{J\left( J+1\right) }\left( \frac{m_{2}}{m_{2}}-\frac{m_{1}%
}{m_{1}}\right) \right)  \notag \\
&&\times \frac{1}{2J+1}\left( Q_{J+1}\left( z\right) -Q_{J-1}\left( z\right)
\right) ,
\end{eqnarray}

Spin-spin interaction

\begin{equation}
\mathcal{K}^{\left( s-s\right) }(p,q)=C_{2}^{2}\frac{\pi q_{1}q_{2}}{%
2m_{1}m_{2}}\left( \frac{p}{q}+\frac{q}{p}\right) Q_{J}\left( z\right)
-C_{2}^{2}\frac{\pi q_{1}q_{2}}{m_{1}m_{2}}\frac{1}{2J+1}\left\{
JQ_{J+1}\left( z\right) +\left( J+1\right) Q_{J-1}\left( z\right) \right\} ,
\end{equation}
where $C_{1}$and $C_{2}$ are defined in section 4.

\vskip0.2truecm

\textbf{The triplet states} $\left( \ell =J-1\ (J\geq 1),\;\ell =J+1\ (J\geq
0)\;P=(-1)^{J}\right) $

From (64), it follows that the kernels $\mathcal{K}_{12}$ and $\mathcal{K}%
_{21}$\ are responsible for mixing of states with $\ell =J-1$ and $\ell =J+1$%
. These kernels have the form 
\begin{equation}
\mathcal{K}_{12}\left( p,q\right) =\mathcal{K}_{21}\left( p,q\right) =\frac{%
\pi q_{1}q_{2}}{5m_{1}m_{2}}\frac{\sqrt{J\left( J+1\right) }}{\left(
2J+1\right) }\left( \frac{p}{q}Q_{J+1}\left( z\right) +\frac{q}{p}%
Q_{J-1}\left( z\right) -2Q_{J}\left( z\right) \right) ,
\end{equation}
where only\ spin-spin interactions contribute. For kernels $\mathcal{K}_{11}$
and $\mathcal{K}_{11}$ we have from (64):

Orbital terms 
\begin{eqnarray}
\mathcal{K}_{11}^{\left( o\right) }\left( p,q\right) &=&\frac{2\pi q_{1}q_{2}%
}{pq}Q_{J-1}\left( z\right) +\frac{\pi q_{1}q_{2}}{2m_{1}m_{2}}\left( \frac{%
m_{1}}{m_{2}}+\frac{m_{2}}{m_{1}}+2-J\right) \left( \frac{p}{q}+\frac{q}{p}%
\right) Q_{J-1}(z)  \notag \\
&&+\frac{\pi q_{1}q_{2}}{m_{1}m_{2}}JQ_{J}\left( z\right) -\frac{\pi
q_{1}q_{2}}{2m_{1}m_{2}}\left( \frac{m_{1}}{m_{2}}+\frac{m_{2}}{m_{1}}%
+2\right) \delta _{J,1},
\end{eqnarray}
\begin{eqnarray}
\mathcal{K}_{22}^{\left( o\right) }\left( p,q\right) &=&\frac{2\pi q_{1}q_{2}%
}{pq}Q_{J+1}\left( z_{1}\right) +\frac{\pi q_{1}q_{2}}{2m_{1}m_{2}}\left( 
\frac{m_{1}}{m_{2}}+\frac{m_{2}}{m_{1}}-J\right) \left( \frac{p}{q}+\frac{q}{%
p}\right) Q_{J+1}(z_{1})  \notag \\
&&+\frac{\pi q_{1}q_{2}}{m_{1}m_{2}}\left( J+2\right) Q_{J+2}\left( z\right)
,
\end{eqnarray}
Spin-orbit interaction 
\begin{equation}
\mathcal{K}_{11}^{\left( s-o\right) }\left( p,q\right) =\frac{\pi q_{1}q_{2}%
}{2m_{1}m_{2}}\frac{J-1}{2J-1}\left( \frac{m_{2}}{m_{1}}+\frac{m_{1}}{m_{2}}%
+4\right) \left( Q_{J}\left( z\right) -Q_{J-2}\left( z\right) \right) ,
\end{equation}
\begin{equation}
\mathcal{K}_{22}^{\left( s-o\right) }\left( p,q\right) =-\frac{\pi q_{1}q_{2}%
}{2m_{1}m_{2}}\left( \frac{m_{2}}{m_{1}}+\frac{m_{1}}{m_{2}}+4\right) \frac{%
J+2}{2J+3}\left( Q_{J+2}\left( z\right) -Q_{J}\left( z\right) \right) ,
\end{equation}
Spin-spin interaction 
\begin{equation}
\mathcal{K}_{11}^{\left( s-s\right) }\left( p,q\right) =\frac{\pi q_{1}q_{2}%
}{2m_{1}m_{2}}\frac{1}{2J+1}\left( \left( \frac{p}{q}+\frac{q}{p}\right)
Q_{J-1}\left( z\right) -2Q_{J}\left( z\right) \right) ,
\end{equation}
\begin{equation}
\mathcal{K}_{22}^{\left( s-s\right) }\left( p,q\right) =\frac{\pi q_{1}q_{2}%
}{2m_{1}m_{2}}\frac{1}{2J+3}\left( \left( \frac{p}{q}+\frac{q}{p}\right)
Q_{J+1}\left( z\right) -2Q_{J+2}\left( z\right) \right) ,
\end{equation}
Annihilation term\textit{\ }(particle-antiparticle system only, with $%
m_{1}=m_{2}=m$, $q_{1}=q_{2}=e$) 
\begin{equation}
\mathcal{K}^{ann}\left( p,q\right) =-\frac{2\pi e^{2}}{m^{2}}\delta _{J,1}.
\end{equation}
We note that in the non-relativistic limit the only terms that survive are
the first terms of the orbital part of the kernels. They have the common
form $2\pi q_{1}q_{2}Q_{\ell }(z)/pq$, where 
\begin{equation}
Q_{\ell }(z)=\frac{pq}{2\pi }\int \frac{d\widehat{\mathbf{q}}d\widehat{%
\mathbf{p}}}{\left( \mathbf{p}-\mathbf{q}\right) ^{2}}Y^{m_{\ell }^{\ast
}}\left( \widehat{\mathbf{p}}\right) Y^{m_{\ell }}\left( \widehat{\mathbf{q}}%
\right) .
\end{equation}
Thus, equations (57), (61) reduce to the form 
\begin{equation}
\left( \omega _{p}+\Omega _{p}-E\right) f^{\ell }(p)=\frac{%
m_{1}m_{2}q_{1}q_{2}}{\pi \sqrt{\omega _{p}\Omega _{p}}p}\int_{0}^{\infty }dq%
\frac{q}{\sqrt{\omega _{q}\Omega _{q}}}Q_{\ell }(z)f^{\ell }(q).
\end{equation}
If we expand the relativistic free-particle energy 
\begin{equation}
\omega _{p}\simeq m_{1}\left( 1+\frac{1}{2}\left( \frac{\mathbf{p}}{m_{1}}%
\right) ^{2}\right) ,\;\;\;\Omega _{p}\simeq m_{2}\left( 1+\frac{1}{2}\left( 
\frac{\mathbf{p}}{m_{2}}\right) ^{2}\right) ,
\end{equation}
we obtain the two particle Schr\"{o}dinger radial equation in momentum space
[18] 
\begin{equation}
\left( \frac{\mathbf{p}^{2}}{2m_{r}}-\varepsilon \right) f^{\ell }(p)=\frac{%
\alpha }{\pi p}\int_{0}^{\infty }dq\,q\,Q_{\ell }(z)f^{\ell }(q),
\end{equation}
where $\alpha =q_{1}q_{2}/4\pi $. The solutions of these equations are well
known. They are given in Appendix D (formula (201)).

\vskip0.8truecm {\normalsize \noindent {\large 7}{\textbf{\large . Energy
eigenvalues and relativistic corrections to }}}$O\left( \alpha ^{4}\right) $ 
{\normalsize {\textbf{\large for arbitrary mass ratio}}}

{\normalsize \vskip 0.4truecm }

The energy eigenvalues $E_{n,J}$ for the states, when $\ell =J$, can be
calculated from the equation 
\begin{eqnarray}
E\int_{0}^{\infty }dp\,p^{2}f^{J}(p)f^{J}(p) &=&\int_{0}^{\infty
}dp\,p^{2}\,\left( \omega _{p}+\Omega _{p}\right) f^{J}(p)f^{J}(p)+  \notag
\\
&&-\frac{m_{1}m_{2}}{\left( 2\pi \right) ^{3}}\int_{0}^{\infty }\frac{dpp^{2}%
}{\sqrt{\omega _{p}\Omega _{p}}}\int_{0}^{\infty }dq\frac{q^{2}}{\sqrt{%
\omega _{q}\Omega _{q}}}\mathcal{K}(p\mathbf{,}q)f^{J}(p)f^{J}(q).
\end{eqnarray}
For the $\ell =J\mp 1$ triplet states we have the matrix equation 
\begin{eqnarray}
E\int_{0}^{\infty }dp\,p^{2}\mathbb{F}^{\dagger }(p)\mathbb{F}(p)
&=&\int_{0}^{\infty }dp\,p^{2}\,\left( \omega _{p}+\Omega _{p}\right) 
\mathbb{F}^{\dagger }(p)\mathbb{F}(p)+  \notag \\
&&-\frac{m_{1}m_{2}}{\left( 2\pi \right) ^{3}}\int_{0}^{\infty }\frac{dpp^{2}%
}{\sqrt{\omega _{p}\Omega _{p}}}\int_{0}^{\infty }dq\frac{q^{2}}{\sqrt{%
\omega _{q}\Omega _{q}}}\mathbb{K}(p\mathbf{,}q)\mathbb{F}^{\dagger }(p)%
\mathbb{F}(q).
\end{eqnarray}
To obtain results for $E$ to $O\left( \alpha ^{4}\right) $ we use the forms
of the kernels expanded to $O\left( p^{2}/m^{2}\right) $ (expressions
(74)-(87)) and replace $f^{\ell }(p)$ by their non-relativistic
(Schr\"{o}dinger) form (201) (see Appendix D). The most important integrals,
which we used for calculating (92) and (93), are given in Appendix D. In
Appendix F we show that the contribution of the kernels $\mathcal{K}_{12}$
and $\mathcal{K}_{21}$ in (94), which are responsible for coupling\ of the
equations (61), is zero. Thus in the framework of the present approximation,
the energy corrections for the triplet states with $\ell =J-1$ and $\ell
=J+1 $ can be calculated independently. We present the results in the form $%
\Delta \varepsilon =E+\frac{\alpha ^{2}m_{r}}{2n^{2}}-M$, $M=m_{1}+m_{2}$, $%
m_{r}=m_{1}m_{2}/\left( m_{1}+m_{2}\right) $.

\vskip0.2truecm

\textbf{Singlet states} $\left( \ell =J=0,\;P=-1\right) $ (which include the
ground state)

The kinetic energy corrections 
\begin{equation}
\Delta \varepsilon _{n}^{(sg)\left( k\right) }=-\frac{\alpha ^{4}m_{r}}{n^{3}%
}\left( 1-\frac{3}{8n}\right) \left( 1-\frac{3m_{r}}{M}\right) ,
\end{equation}
Orbital energy corrections 
\begin{equation}
\Delta \varepsilon _{n}^{(sg)\left( o\right) }=-\frac{\alpha ^{4}m_{r}}{n^{3}%
}\left( -\frac{1}{2}+\frac{3m_{r}}{M}-\frac{m_{r}}{M}\frac{1}{n}\right) ,
\end{equation}
Spin-orbit energy corrections 
\begin{equation}
\Delta \varepsilon _{n}^{(sg)\left( s-o\right) }=0,
\end{equation}
Spin-spin energy corrections 
\begin{equation}
\Delta \varepsilon _{n}^{(sg)\left( s-s\right) }=-\frac{\alpha ^{4}m_{r}}{%
n^{3}}\frac{2m_{r}}{M}.
\end{equation}
The total energy corrections 
\begin{equation}
\Delta \varepsilon _{n}^{(sg)}=-\frac{\alpha ^{4}m_{r}}{n^{3}}\left( \frac{1%
}{2}+\frac{2m_{r}}{M}-\frac{1}{8n}\left( 3-\frac{m_{r}}{M}\right) \right) .
\end{equation}
\vskip0.2truecm

\textbf{Quasi-singlet, quasi-triplet states} $\left( \ell =J\ (J\geq
1),\;P=(-1)^{J+1}\right) $

The kinetic energy and orbital energy corrections are common for
quasi-singlet and quasi-triplet states: 
\begin{equation}
\Delta \varepsilon _{n,J}^{\left( k\right) }=-\frac{\alpha ^{4}m_{r}}{2n^{3}}%
\left( 1-\frac{3m_{r}}{M}\right) \left( \frac{2}{2J+1}-\frac{3}{4n}\right) ,
\end{equation}
Orbital energy corrections 
\begin{equation}
\Delta \varepsilon _{n,J}^{(orb)}=-\frac{\alpha ^{4}m_{r}}{n^{3}}\frac{m_{r}%
}{M}\left( \frac{3}{2J+1}-\frac{1}{n}\right) ,
\end{equation}
Spin-orbit energy corrections 
\begin{equation}
\Delta \varepsilon _{n,J}^{(s-o)}=\frac{\alpha ^{4}m_{r}}{2n^{3}}\frac{1}{%
M^{2}}\frac{-C_{2}^{2}\left( \left( m_{1}+m_{2}\right)
^{2}+2m_{1}m_{2}\right) +2C_{1}C_{2}\sqrt{J\left( J+1\right) }\left(
m_{2}^{2}-m_{1}^{2}\right) }{\left( 2J+1\right) \left( J+1\right) J},
\end{equation}
Spin-spin energy corrections 
\begin{equation}
\Delta \varepsilon _{n,J}^{(s-s)}=\frac{\alpha ^{4}m_{r}}{n^{3}}\frac{m_{r}}{%
M}\frac{C_{2}^{2}}{\left( 2J+1\right) \left( J+1\right) J}.
\end{equation}
The coefficients $C_{1}$ and $C_{2}$ are defined here in the same way as in
formulas (78), (79).

Thus, the total energy corrections are 
\begin{equation}
\Delta \varepsilon _{n,J,s}=-\frac{\alpha ^{4}m_{r}}{n^{3}}\left( \frac{1}{%
2J+1}\left( 1+\frac{1\mp \xi ^{-1}}{4J\left( J+1\right) }\right) -\frac{1}{8n%
}\left( 3-\frac{m_{r}}{M}\right) \right) ,
\end{equation}
where upper and lower signs correspond to quasi-singlet $s=s_{s}=0$ and
quasi-triplet $s=s_{t}=1$ states respectively.

\vskip0.2truecm

\textbf{Triplet states} $\left( \ell =J-1\ (J\geq 1),\;P=(-1)^{J}\right) $

Kinetic energy corrections 
\begin{equation}
\Delta \varepsilon _{n,J}^{(tr)(k)}=-\frac{\alpha ^{4}m_{r}}{n^{3}}\left( 
\frac{1}{2J-1}-\frac{3}{8n}\right) \left( 1-\frac{3m_{r}}{M}\right) ,
\end{equation}
Orbital energy corrections 
\begin{equation}
\Delta \varepsilon _{n,J}^{\left( tr\right) \left( o\right) }=-\frac{\alpha
^{4}m_{r}}{n^{3}}\left( \frac{3}{2J-1}-\frac{\delta _{J,1}}{2}-\frac{1}{n}%
\right) \frac{m_{r}}{M},
\end{equation}
Spin-orbit energy corrections 
\begin{equation}
\Delta \varepsilon _{n,J}^{(tr)\left( s-o\right) }=\frac{\alpha ^{4}m_{r}}{%
n^{3}}\frac{1-\delta _{J,1}}{2J\left( 2J-1\right) }\left( 1+\frac{2m_{r}}{M}%
\right) ,
\end{equation}
Spin-spin energy corrections 
\begin{equation}
\Delta \varepsilon _{n,J}^{(tr)\left( s-s\right) }=-\frac{\alpha ^{4}m_{r}}{%
n^{3}}\left( \frac{1-\delta _{J,1}}{J\left( 2J+1\right) \left( 2J-1\right) }-%
\frac{2}{3}\delta _{J,1}\right) \frac{m_{r}}{M}.
\end{equation}
The total energy corrections 
\begin{equation}
\Delta \varepsilon _{n,J}^{tr}=-\frac{\alpha ^{4}m_{r}}{n^{3}}\left( \frac{1%
}{2J-1}\left( 1-\frac{1}{2J}-\frac{2m_{r}}{M}\frac{1}{2J+1}\right) -\frac{1}{%
8n}\left( 3-\frac{m_{r}}{M}\right) \right) .
\end{equation}
The annihilation term for positronium-like systems is 
\begin{equation}
\Delta \varepsilon _{n}^{\left( anh\right) }=\frac{\alpha ^{4}m}{4n^{3}}%
\delta _{J,1}.
\end{equation}

\textbf{Triplet states }$\left( \ell =J+1\ (J\geq 0),\;\;P=(-1)^{J}\right) $

The kinetic energy corrections 
\begin{equation}
\Delta \varepsilon _{n,J}^{(tr)(k)}=-\frac{\alpha ^{4}m_{r}}{n^{3}}\left( 
\frac{1}{2J+3}-\frac{3}{8n}\right) \left( 1-\frac{3m_{r}}{M}\right) ,
\end{equation}
Orbital energy corrections 
\begin{equation}
\Delta \varepsilon _{n,J}^{\left( tr\right) \left( o\right) }=-\frac{\alpha
^{4}m_{r}}{n^{3}}\left( \frac{3}{2J+3}-\frac{1}{n}\right) \frac{m_{r}}{M},
\end{equation}
Spin-orbit energy corrections 
\begin{equation}
\Delta \varepsilon _{n,J}^{(tr)\left( s-o\right) }=-\frac{\alpha ^{4}m_{r}}{%
n^{3}}\frac{1}{2\left( J+1\right) \left( 2J+3\right) }\left( 1+\frac{2m_{r}}{%
M}\right) ,
\end{equation}
Spin-spin energy corrections 
\begin{equation}
\Delta \varepsilon _{n,J}^{(tr)\left( s-s\right) }=-\frac{\alpha ^{4}m_{r}}{%
n^{3}}\frac{1}{\left( J+1\right) \left( 2J+3\right) \left( 2J+1\right) }%
\frac{m_{r}}{M}.
\end{equation}
The total energy corrections 
\begin{equation}
\Delta \varepsilon _{n,J}^{tr}=-\frac{\alpha ^{4}m_{r}}{n^{3}}\left( \frac{1%
}{2J+3}\left( 1+\frac{1}{2\left( J+1\right) }+\frac{2m_{r}}{M}\frac{1}{2J+1}%
\right) -\frac{1}{8n}\left( 3-\frac{m_{r}}{M}\right) \right) .
\end{equation}

For two equal masses our calculations agree with positronium\ results [19].
In the limit when one of the masses becomes infinite, say $m_{2}\rightarrow
\infty $, the above results reduce to those obtained for a one-electron
Dirac equation in a static Coulomb potential (to $O\left( \alpha ^{4}\right) 
$), namely 
\begin{equation}
\Delta \varepsilon _{n,j}=-\frac{\alpha ^{4}m_{1}}{n^{3}}\left( \frac{1}{2j+1%
}-\frac{3}{8n}\right) .
\end{equation}
Indeed, when $m_{2}\rightarrow \infty $ not only the total angular momentum,
but also $\mathbf{J}_{1}^{2}=\left( \mathbf{L}_{1}+\mathbf{S}_{1}\right)
^{2} $ and $\mathbf{S}_{2}^{2}$ are independently conserved. Thus, in this
case, we can replace the quantum number $J$ by $j-1/2$\ in (99)-(103)(when $%
s=0$) and (110)-(114), and by $j+1/2$ in expressions (99)-(103)(when $s=1$)
and (104)-(108). In other words 
\begin{eqnarray}
J &\rightarrow &j-1/2\;\ \ \ \ \ for\;\ \ \mid sg_{q}\rangle \;\ \ \ and\;\
\ \ \mid tr_{\ell =J+1}\rangle , \\
J &\rightarrow &j+1/2\;\ \ \ \ \ for\;\ \ \mid tr_{q}\rangle \;\ \ \ and\;\
\ \ \ \mid tr_{\ell =J-1}\rangle .  \notag
\end{eqnarray}
The quantum number $j=\left| \ell \pm 1/2\right| $ belongs here to the
particle with mass $m_{1}$.

The results (98), (103), (108) and (114) agree with the calculations of
Connell, based on a quasipotential reduction of the Bethe-Salpeter equation
[20],\ those of Hersbach, who used a relativistic Lippman-Schwinger
formulation [21], and those of Duviryak and Darewych based on a two-fermion
Breit equation [22]. Grandy [8] obtained the same results from a
perturbative solution of the coupled, nonlinear Dirac equations (8) and (9).

\vskip0.8truecm {\normalsize \noindent {\large 8}{\textbf{\large . Fine and
hyperfine structure. Recoil effects.}} }

{\normalsize \vskip0.4truecm }

In this section we shall analyze the formulas obtained in the previous
section and we shall apply them to the energy spectra of some exotic atoms.
To compare our calculations with experimental data we shall use the standard
spectroscopical notation. First of all, it follows from (94)-(96) and
(104)-(106) that the difference between energy levels of $2S$ and $1S$
states is given by formula 
\begin{equation}
E\left( 2S\right) -E\left( 1S\right) =\frac{3\alpha ^{2}m_{r}}{8}+\frac{%
\alpha ^{4}m_{r}}{128}\left( 11+15\frac{m_{r}}{M}\right) .
\end{equation}
Note that the formula (117) ignores hyperfine splitting, that is we exclude
the spin-spin interaction. The fine structure of $2P$-state follows
similarly from (104)-(106) and (110)-(112), provided that we exclude the
spin-spin interaction (107) and (113) 
\begin{equation}
\Delta E_{fs}\left( 2P\right) \equiv E\left( 2P_{3/2}\right) -E\left(
2P_{1/2}\right) =\frac{\alpha ^{4}m_{r}}{32}\left( 1+\frac{2m_{r}}{M}\right)
.
\end{equation}
The HFS of $1S_{1/2}$ and $2S_{1/2}$\ is obtained from (98) and (108): 
\begin{equation}
\Delta E_{hfs}\left( 1S_{1/2}\right) =\alpha ^{4}m_{r}\frac{8m_{r}}{3M},
\end{equation}
\begin{equation}
\Delta E_{hfs}\left( 2S_{1/2}\right) =\alpha ^{4}m_{r}\frac{m_{r}}{3M}.
\end{equation}
Actually this hyperfine splitting, expressions (119) and (120), arises from
the difference of spin-spin terms (97) and (107).\ The formulae (119) and
(120) give the usual Fermi splitting [18].

The HFS of states with $\ell >0$ is more complicated. For each $\ell >0$\ we
have two sorts of states with different $J$ and $s$: states with $J=\ell +1$
and $J=\ell $, $s=s_{s}$ and states with $J=\ell -1$ and $J=\ell $, $s=s_{t}$%
. From (103),(108) and (114) we obtain the general HFS formulae for all
quantum numbers $n$, $\ell $ and for any mass ratio. 
\begin{equation}
\Delta E_{hfs}\left( n,\ell ,s_{s}\right) \equiv \Delta \varepsilon
_{n,J=\ell +1}^{tr}-\Delta \varepsilon _{n,J=\ell ,s_{s}}=\frac{\alpha
^{4}m_{r}}{n^{3}}\frac{1}{2\ell +1}\left( \frac{2\ell +1-\xi ^{-1}}{4\ell
\left( \ell +1\right) }+\frac{2m_{r}}{M}\frac{1}{2\ell +3}\right) ,
\end{equation}
\begin{equation}
\Delta E_{hfs}\left( n,\ell ,s_{t}\right) \equiv \Delta \varepsilon
_{n,J=\ell ,s_{t}}-\Delta \varepsilon _{n,J=\ell -1}^{tr}=\frac{\alpha
^{4}m_{r}}{n^{3}}\frac{1}{2\ell +1}\left( \frac{2\ell +1-\xi ^{-1}}{4\ell
\left( \ell +1\right) }+\frac{2m_{r}}{M}\frac{1}{2\ell -1}\right) .
\end{equation}
The quantity $\xi $ is defined by (53), but with the quantum number $J$
replaced by $\ell $.

For the particular case when $m_{2}>>m_{1}$ we obtain from (121) and (122) 
\begin{equation}
\Delta E_{hfs}\left( n,\ell ,s_{s}\right) =\frac{\alpha ^{4}m_{1}}{n^{3}}%
\frac{8\left( \ell +1\right) }{\left( 2\ell +1\right) ^{2}\left( 2\ell
+3\right) }\frac{m_{1}}{m_{2}},
\end{equation}
\begin{equation}
\Delta E_{hfs}\left( n,\ell ,s_{t}\right) =\frac{\alpha ^{4}m_{1}}{n^{3}}%
\frac{8\ell }{\left( 2\ell +1\right) ^{2}\left( 2\ell -1\right) }\frac{m_{1}%
}{m_{2}},
\end{equation}
to leading order in $m_{1}/m_{2}$.

The approximate results (123) and (124) are the HFS formulae usually quoted
in the literature (e.g. [18] p. 110, and [24] p. 836). As mentioned in
section 7, in the one-body limit ($m_{2}\rightarrow \infty $), states with
labels $s_{s}$ and $s_{t}$\ can be characterized by the quantum numbers $%
j=\ell +1/2$ and $j=\ell -1/2$ respectively. In this one-body picture, the
labels $s_{s}$ and $s_{t}$ should be replaced by $j$ in the formulas (121),
(124). Note here that, because of the mixed nature of the states $n\ell
_{\ell -1/2}\left( \ell =J,\;s_{t}=1\right) $ and $n\ell _{\ell +1/2}\left(
\ell =J,\;s_{s}=0\right) $, the spin-orbit interaction contributes to the
spliting as well. The splitting is largest when $m_{1}=m_{2}$ and disappears
for all states in the one-body limit.

It is easy to get the HFS of $2P_{1/2},\;2P_{3/2}$ states from (121) and
(122): 
\begin{equation}
\Delta E_{hfs}\left( 2P_{1/2}\right) =\alpha ^{4}m_{r}\left( \frac{1}{64}-%
\frac{\xi ^{-1}}{192}+\frac{1}{12}\frac{m_{r}}{M}\right) ,
\end{equation}
\begin{equation}
\Delta E_{hfs}\left( 2P_{3/2}\right) =\alpha ^{4}m_{r}\left( \frac{1}{64}-%
\frac{\xi ^{-1}}{192}+\frac{1}{60}\frac{m_{r}}{M}\right) ,
\end{equation}
where $\xi ^{-1}=\sqrt{8\left( \frac{m_{1}-m_{2}}{M}\right) ^{2}+1}$. Note
that, if we use the pure triplet state instead of the mixed states, we get
the following expressions for hyperfine splitting: $E_{hfs}\left(
2P_{1/2}\right) =\alpha ^{4}m_{r}\left( \frac{1}{96}+\frac{m_{r}}{12M}%
\right) $, $E_{hfs}\left( 2P_{3/2}\right) =\alpha ^{4}m_{r}\left( \frac{1}{96%
}+\frac{m_{r}}{60M}\right) .$ These give the correct results only for $%
m_{1}=m_{2}$ (positronium).

The results of our calculations and comparison with experimental data for
muonium, hydrogen and muonic hydrogen are summarized in Tables 2-4. In each
table the first two entries reflect, primary, fine-structure splittings,
while $\Delta E_{L}$ corresponds to the $O\left( \alpha ^{4}\right) $ recoil
contribution to the Lamb shift, $E\left( 2S_{1/2}\right) -E(2P_{1/2})$.\
Note that except for the hyperfine splitting of the ground state, which has
been measured very accurately, the experimental results are quoted in the
tables to the number of significant digits available. The calculated
results, we must remember, are accurate to $O\left( \alpha ^{4}\right) $
only, where as the experimental results are exact except for experimental
error, that is they contain all orders in $\alpha $.\ 

Although hydrogen and muonic hydrogen are not strictly leptonic atoms, we
can take into account the proton anomalous magnetic moment in the hyperfine
splitting in\ the same phenomenological manner as was done in [23]. Thus, to
get the corrected result for HFS in hydrogen and muonic hydrogen, we
multiply the expression for $\Delta E_{hfs}$ by the factor $\left(
1+k_{p}\right) $, where $k_{p}=1.792847$ is the\ anomalous magnetic moment
of the proton. The last columns in tables 3, 4 give the HFS corrected by the
factor $\left( 1+k_{p}\right) $. The anomalous magnetic moment of the muon
is very small, nevertheless for the sake of completeness, we provide the
same corrections for muonium, where we use the factor $\left( 1+a_{\mu
}\right) $ ($a_{\mu }=0.001166$) to get the result of the last column in
Table 2 from those of the third column. We note that doing so improves the
agrement between the calculated and the experimental HFS results in most
cases, and dramatically so for hydrogen and muonic hydrogen.

{\normalsize \vskip0.4truecm }

\begin{center}
Table 2. Energy shifts, fine and hyperfine structure in muonium for $n=1,2.$

\begin{tabular}{||c|r|r|r||}
\hline
$\mu ^{+}e^{-}$ & experiment [24] & theory & $a_{\mu }$-correction \\ \hline
$E\left( 2S_{1/2}\right) -E(1S_{1/2})$ & $2455.3\;THz$ & $2455.54\;THz$ & -
\\ \hline
$\Delta E_{fs}\left( 2P\right) $ & $10922\;MHz$ & $11000.85\;MHz$ & - \\ 
\hline
$\Delta E_{L}$ & $1047\;MHz$ & $69.59\;MHz$ & - \\ \hline
$\Delta E_{hfs}\left( 1S_{1/2}\right) $ & $4463\;MHz$ & $4453.84\;MHz$ & $%
4459.03\;MHz$ \\ \hline
$\Delta E_{hfs}\left( 2S_{1/2}\right) $ & $558\;MHz$ & $556.73\;MHz$ & $%
557.38\;MHz$ \\ \hline
$\Delta E_{hfs}\left( 2P_{1/2}\right) $ & $186\;MHz$ & $185.80\;MHz$ & $%
186.01\;MHz$ \\ \hline
$\Delta E_{hfs}\left( 2P_{3/2}\right) $ & $74\;MHz$ & $74.43\;MHz$ & $%
74.52\;MHz$ \\ \hline
\end{tabular}
\end{center}

{\normalsize \vskip0.4truecm }

\begin{center}
Table 3. Energy shifts, fine and hyperfine structure in hydrogen for $%
n=1,\;2 $.

\begin{tabular}{||c|r|r|r||}
\hline
$H$ & experiment [25], [26] & theory & $k_{p}$-correction \\ \hline
$E\left( 2S_{1/2}\right) -E(1S_{1/2})$ & $2466\;THz$ [25] & $2469.99\;THz$ & 
- \\ \hline
$\Delta E_{fs}\left( 2P\right) $ & $10969.04\;MHz$ [25] & $10954.77\;MHz$ & -
\\ \hline
$\Delta E_{L}$ & $1057.85\;MHz$ [25] & $7.94\;MHz$ & - \\ \hline
$\Delta E_{hfs}\left( 1S_{1/2}\right) $ & $1420.4058\;MHz$ [26] & $%
508.0265\;MHz$ & $1418.8403\;MHz$ \\ \hline
$\Delta E_{hfs}\left( 2S_{1/2}\right) $ & $177.5569\;MHz$ [25] & $%
63.5033\;MHz$ & $177.3550\;MHz$ \\ \hline
$\Delta E_{hfs}\left( 2P_{1/2}\right) $ & $59.1721\;MHz$ [25] & $%
21.1698\;MHz $ & $59.1241\;MHz$ \\ \hline
$\Delta E_{hfs}\left( 2P_{3/2}\right) $ & $23.6541\;MHz$ [25] & $8.4696\;MHz$
& $23.6543\;MHz$ \\ \hline
\end{tabular}
\end{center}

{\normalsize \vskip0.4truecm }

\begin{center}
Table 4. Energy shifts, fine and hyperfine structure in muonic hydrogen for $%
n=1,\;2$.

\begin{tabular}{||c|r|r|r||}
\hline
$\mu ^{-}H$ & theory [27], experiment [28] & theory & $k_{p}$-correction \\ 
\hline
$E\left( 2S_{1/2}\right) -E(1S_{1/2})$ & - & $1.90\;eV$ & - \\ \hline
$\Delta E_{fs}\left( 2P\right) $ & $8.35\;meV\;$[27] & $9.95\;meV$ & - \\ 
\hline
$\Delta E_{L}$ & $-202.07\;meV\;$[27] & $+1.02\;meV$ & - \\ \hline
$\Delta E_{hfs}\left( 1S_{1/2}\right) $ & $183\;meV\;$[28] & $65.33\;meV$ & $%
142.36\;meV$ \\ \hline
$\Delta E_{hfs}\left( 2S_{1/2}\right) $ & $22.75\;meV\;$[27] & $8.17\;meV$ & 
$22.81\;meV$ \\ \hline
$\Delta E_{hfs}\left( 2P_{1/2}\right) $ & $7.96\;meV\;$[27] & $2.79\;meV$ & $%
7.79\;meV$ \\ \hline
$\Delta E_{hfs}\left( 2P_{3/2}\right) $ & $3.39\;meV\;$[27] & $1.16\;meV$ & $%
3.23\;meV$ \\ \hline
\end{tabular}
\end{center}

From tables 2-4 we see that the $O\left( \alpha ^{4}\right) $ calculated
results agree with observation remarkably well, except, of course, for the
Lamb shift, which is dominated by corrections of higher-order that $\alpha
^{4}$. The present theoretical results for $\Delta E_{L}$ reflect only the $%
O\left( \alpha ^{4}\right) $ recoil effects, and these provide a small,
though not insignificant, contribution. Note that if we use the approximate $%
m_{1}/m_{2}$ expansion formulas (123) and (124), for example for HFS of $%
2P_{1/2}$ and $2P_{3/2}$ states we get the result (with $a_{\mu }$ and $%
k_{p} $ corrections) $\Delta E_{hfs}\left( 2P_{1/2}\right) =188.50\;MHz$, $%
\Delta E_{hfs}\left( 2P_{3/2}\right) =75.40\;MHz$ for muonium, and $\Delta
E_{hfs}\left( 2P_{1/2}\right) =10.47\;meV$, $\Delta E_{hfs}\left(
2P_{3/2}\right) =4.19\;meV$\ for muonic hydrogen. These approximate results
do not agree as well with observation as the values calculated without the $%
m_{1}/m_{2}$ expansion, and hence they differ from what we give in last
columns of tables 2 and 4. Note that this difference can be of the same
order as the contribution of higher order in $\alpha $ for HFS. The $%
m_{1}/m_{2}$ expansion approximation\ is not significant for hydrogen for
which the mass ratio is very small, but it is appreciable for muonium and
particularly for muonic hydrogen.

Since our results are true for arbitrary mass ratios we can speak here about
recoil \textit{effects} rather than about\ recoil \textit{corrections}. The
recoil effect contributes to the Lamb shift (shift between $2S_{1/2}$ and $%
2P_{1/2}$ levels). However the states $2S_{1/2}$ and $2P_{1/2}$ are split
(hyperfine splitting as discussed below) because of the spin-spin
interaction, so to calculate the Lamb shift we have to take the energy
corrections for each state without spin-spin terms. For the state $2S_{1/2}$%
\ it is the sum of the energy corrections (94) and (95), while for the $%
2P_{1/2}$ state we take the sum of (110), (111), and (112). Thus, the
formula for the $O\left( \alpha ^{4}\right) $ contribution to the Lamb shift
produced by the recoil effect is 
\begin{equation}
\Delta E_{L}^{\left( rcl\right) }\equiv E\left( 2S_{1/2}\right) -E\left(
2P_{1/2}\right) =\frac{\alpha ^{4}m_{r}}{24}\frac{m_{r}}{M}.
\end{equation}
It is easy to see that the shift (127) is largest when $m_{1}=m_{2}$, and
disappears in the one-body limit, when $m_{2}\rightarrow \infty $.

There is a well known formula for the recoil correction of order $\alpha
^{4} $ obtained on the basis of the Breit interaction [29], [30] 
\begin{equation}
\Delta \varepsilon _{n,\ell ,j}=\frac{\alpha ^{4}m_{r}}{2n^{3}}\frac{%
m_{r}^{2}}{m_{2}^{2}}\left( \frac{1}{j+1/2}-\frac{1}{\ell +1/2}\right)
\left( 1-\delta _{\ell ,0}\right) ,
\end{equation}
which predicts the following contribution to the Lamb shift: 
\begin{equation}
\Delta E_{L}^{\left( rcl\right) }=\Delta \varepsilon _{2,0,1/2}-\Delta
\varepsilon _{2,1,1/2}=-\frac{\alpha ^{4}m_{r}}{48}\frac{m_{r}^{2}}{m_{2}^{2}%
}.
\end{equation}
Comparison of the\ formulas (127) and (129) shows that the total recoil
effect (127) is larger in magnitude than (129), and has opposite sign.\ The
difference between (127) and (129) stems from the spin-orbit interaction: If
we ignore the term $\left( \frac{m_{1}}{m_{2}}+2\right) \sigma _{2}$ in
expression (70), we get instead of (106) and (112) the following result for
the spin-orbit contribution:

for $\ell =J-1$ 
\begin{equation}
\Delta \varepsilon _{n,J}^{(tr)\left( s-o\right) }=\frac{\alpha ^{4}m_{r}}{%
n^{3}}\frac{1-\delta _{J,1}}{2J\left( 2J-1\right) }\left( 1-\frac{m_{r}^{2}}{%
m_{2}^{2}}\right) ,
\end{equation}

for $\ell =J+1$ 
\begin{equation}
\Delta \varepsilon _{n,J}^{(tr)\left( s-o\right) }=-\frac{\alpha ^{4}m_{r}}{%
n^{3}}\frac{1}{2\left( J+1\right) \left( 2J+3\right) }\left( 1-\frac{%
m_{r}^{2}}{m_{2}^{2}}\right) .
\end{equation}
The relevant terms are the second ones in (130), (131). We can replace the ``%
$J$''-quantum numbers in (130) and (131) using (116): 
\begin{equation}
\frac{1}{J\left( 2J-1\right) }=-\left( \frac{1}{J}-\frac{2}{2J-1}\right)
\rightarrow -\left( \frac{1}{j+1/2}-\frac{1}{\ell +1/2}\right) ,
\end{equation}
and 
\begin{equation}
\frac{1}{\left( J+1\right) \left( 2J+3\right) }=\frac{1}{J+1}-\frac{2}{2J+3}%
\rightarrow \frac{1}{j+1/2}-\frac{1}{\ell +1/2}.
\end{equation}
It is now easy to see that the second terms in (130) and (131) give the
formula (128) which, therefore, must be regarded as incomplete. As already
mentioned,\ the proper recoil contribution to the Lamb shift of muonium,
hydrogen and muonic hydrogen, based on our result (127), is given in Tables
2-4.

\vskip0.8truecm {\normalsize \noindent {\large 9}{\textbf{\large .
Concluding remarks}} }

{\normalsize \vskip0.4truecm }

{\normalsize 
}We have used the variational method within the Hamiltonian formalism of QED
to derive relativistic momentum-space wave equations for two-fermion\
systems like muonium. The trial states are chosen to be eigenstates of the
total angular momentum operators $\widehat{\mathbf{J}}^{2}$, $\widehat{J}%
_{3} $ and parity, as well as charge conjugation for particle-antiparticle
systems. A general relativistic reduction of the wave equations to radial
form is given for arbitrary masses of the two fermions. For given $J$ there
is a single radial equation for total spin zero singlet states, but for spin
triplet states there are, in general coupled equations. We have shown how
classification of the states follows naturally from the system of eigenvalue
equations (36), given our trial state.

It is not possible, as far as we know, to obtain analytic solutions of our
relativistic radial equations nor the resulting eigenvalues of the two
fermion system that they describe. However, it is possible to obtain $%
O(\alpha ^{4})$ corrections to the energy eigenvalues analytically for all
states using perturbation theory.

We have compared our calculated results with experiment for fine and
hyperfine splitting of low-lying levels in muonium, hydrogen and muonic
hydrogen. We find good agreement for muonium, as well as for hydrogen and
muonic hydrogen provided that we take into account the anomalous magnetic
moment of the proton in the later two cases.

The method presented here can be generalized to include effects higher order
in alpha by using dressed propagators in place of the bare propagators. This
shall be the subject of a forthcoming work.

{\normalsize 
}

\vskip0.8truecm {\normalsize \noindent {Acknowledgment} }

{\normalsize \vskip0.4truecm }

{\normalsize 
}

The financial support of the National Sciences and Engineering Research
Council of Canada for this work is gratefully acknowledged.

{\normalsize 
} \newpage {\normalsize \noindent {\textbf{\large Appendix A: Total angular
momentum operator in relativistic form}} }

{\normalsize \vskip0.4truecm }

{\normalsize 
}

The total angular momentum operator is defined by the expression 
\begin{equation}
\widehat{\mathbf{J}}=\int d^{3}\mathbf{x\,}\psi ^{\dagger }\left( x\right) 
\mathbf{(}\widehat{\mathbf{L}}+\widehat{\mathbf{S}})\psi \left( x\right)
+\int d^{3}\mathbf{x\,}\phi ^{\dagger }\left( x\right) \mathbf{(}\widehat{%
\mathbf{L}}+\widehat{\mathbf{S}})\phi \left( x\right) ,
\end{equation}
where $\widehat{\mathbf{L}}$ is the orbital angular momentum and $\widehat{%
\mathbf{S}}$ - the spin operator: $\widehat{\mathbf{L}}=\widehat{\mathbf{x}}%
\times \widehat{\mathbf{p}}$ and $\widehat{\mathbf{S}}=\frac{1}{2}\widehat{%
\overrightarrow{\sigma }}$. We use the standard representation for the Pauli
matrices 
\begin{equation}
\widehat{\overrightarrow{\sigma }}=\left[ 
\begin{array}{cc}
\overrightarrow{\sigma } & 0 \\ 
0 & \overrightarrow{\sigma }
\end{array}
\right] ,
\end{equation}
\begin{equation}
\sigma _{1}=\left[ 
\begin{array}{cc}
0 & 1 \\ 
1 & 0
\end{array}
\right] ,\;\;\ \ \ \sigma _{2}=\left[ 
\begin{array}{cc}
0 & -i \\ 
i & 0
\end{array}
\right] ,\;\;\;\sigma _{3}=\left[ 
\begin{array}{cc}
1 & 0 \\ 
0 & -1
\end{array}
\right] .
\end{equation}
Using the field operators $\psi \left( x\right) $ and $\phi \left( x\right) $
in the form (17), (18), after tedious calculations\ we obtain the expression
for operator $\widehat{\mathbf{J}}$. It consists of three parts: two total
angular momentum operators of each particle-antiparticle system [10] and the
following part, relevant to our considerations 
\begin{equation*}
\widehat{J}_{1}=\int d^{3}\mathbf{q}\left( 
\begin{array}{c}
\widehat{L}_{q1}\left( b_{\mathbf{q}\uparrow }^{\dagger }b_{\mathbf{q}%
\uparrow }+b_{\mathbf{q}\downarrow }^{\dagger }b_{\mathbf{q}\downarrow }+D_{%
\mathbf{q}\uparrow }^{\dagger }D_{\mathbf{q}\uparrow }+D_{\mathbf{q}%
\downarrow }^{\dagger }D_{\mathbf{q}\downarrow }\right) \\ 
+\frac{1}{2}\left( b_{\mathbf{q}\uparrow }^{\dagger }b_{\mathbf{q}\downarrow
}+b_{\mathbf{q}\downarrow }^{\dagger }b_{\mathbf{q}\uparrow }+D_{\mathbf{q}%
\downarrow }^{\dagger }D_{\mathbf{q}\uparrow }+D_{\mathbf{q}\uparrow
}^{\dagger }D_{\mathbf{q}\downarrow }\right)
\end{array}
\right) ,
\end{equation*}
\begin{equation}
\widehat{J}_{2}=\int d^{3}\mathbf{q}\left( 
\begin{array}{c}
\widehat{L}_{q2}\left( b_{\mathbf{q}\uparrow }^{\dagger }b_{\mathbf{q}%
\uparrow }+b_{\mathbf{q}\downarrow }^{\dagger }b_{\mathbf{q}\downarrow }+D_{%
\mathbf{q}\uparrow }^{\dagger }D_{\mathbf{q}\uparrow }+D_{\mathbf{q}%
\downarrow }^{\dagger }D_{\mathbf{q}\downarrow }\right) \\ 
+\frac{i}{2}\left( -b_{\mathbf{q}\uparrow }^{\dagger }b_{\mathbf{q}%
\downarrow }+b_{\mathbf{q}\downarrow }^{\dagger }b_{\mathbf{q}\uparrow }-D_{%
\mathbf{q}\uparrow }^{\dagger }D_{\mathbf{q}\downarrow }+D_{\mathbf{q}%
\downarrow }^{\dagger }D_{\mathbf{q}\uparrow }\right)
\end{array}
\right) ,
\end{equation}
\begin{equation*}
\widehat{J}_{3}=\int d^{3}\mathbf{q}\left( 
\begin{array}{c}
\widehat{L}_{q3}\left( b_{\mathbf{q}\uparrow }^{\dagger }b_{\mathbf{q}%
\uparrow }+b_{\mathbf{q}\downarrow }^{\dagger }b_{\mathbf{q}\downarrow }+D_{%
\mathbf{q}\uparrow }^{\dagger }D_{\mathbf{q}\uparrow }+D_{\mathbf{q}%
\downarrow }^{\dagger }D_{\mathbf{q}\downarrow }\right) \\ 
+\frac{1}{2}\left( b_{\mathbf{q}\uparrow }^{\dagger }b_{\mathbf{q}\uparrow
}-b_{\mathbf{q}\downarrow }^{\dagger }b_{\mathbf{q}\downarrow }+D_{\mathbf{q}%
\uparrow }^{\dagger }D_{\mathbf{q}\uparrow }-D_{\mathbf{q}\downarrow
}^{\dagger }D_{\mathbf{q}\downarrow }\right)
\end{array}
\right) .
\end{equation*}
Here $\widehat{\mathbf{L}}_{q}$ is the orbital angular momentum operator in
momentum representation:

\begin{equation}
(\widehat{\mathbf{L}}_{q})_{i}\equiv \widehat{L}_{qi}=-i\left( \mathbf{%
q\times }\nabla _{q}\right) _{i}.
\end{equation}
Note that these expressions are valid for any $t$, since the time-dependent
phase factors of the form $e^{i\omega _{q}t}$\ cancel out. For the operator $%
\widehat{\mathbf{J}}^{2}=\widehat{J}_{1}^{2}+\widehat{J}_{2}^{2}+\widehat{J}%
_{3}^{2}$ we have 
\begin{eqnarray}
\widehat{\mathbf{J}}^{2} &=&\int d^{3}\mathbf{q}\left( 
\begin{array}{c}
\left( \widehat{\mathbf{L}}_{q}^{2}+\frac{3}{4}\right) \left( b_{\mathbf{q}%
\uparrow }^{\dagger }b_{\mathbf{q}\uparrow }+b_{\mathbf{q}\downarrow
}^{\dagger }b_{\mathbf{q}\downarrow }+D_{\mathbf{q}\uparrow }^{\dagger }D_{%
\mathbf{q}\uparrow }+D_{\mathbf{q}\downarrow }^{\dagger }D_{\mathbf{q}%
\downarrow }\right) \\ 
+\widehat{L}_{q-}b_{\mathbf{q}\uparrow }^{\dagger }b_{\mathbf{q}\downarrow }+%
\overset{\symbol{94}}{L}_{q+}b_{\mathbf{q}\downarrow }^{\dagger }b_{\mathbf{q%
}\uparrow }+\overset{\symbol{94}}{L}_{q-}D_{\mathbf{q}\uparrow }^{\dagger
}D_{\mathbf{q}\downarrow }+\overset{\symbol{94}}{L}_{q+}D_{\mathbf{q}%
\downarrow }^{\dagger }D_{\mathbf{q}\uparrow } \\ 
+\widehat{L}_{q3}\left( b_{\mathbf{q}\uparrow }^{\dagger }b_{\mathbf{q}%
\uparrow }-b_{\mathbf{q}\downarrow }^{\dagger }b_{\mathbf{q}\downarrow }+D_{%
\mathbf{q}\uparrow }^{\dagger }D_{\mathbf{q}\uparrow }-D_{\mathbf{q}%
\downarrow }^{\dagger }D_{\mathbf{q}\downarrow }\right)
\end{array}
\right)  \notag \\
&&+\frac{1}{2}\int d^{3}\mathbf{q}^{\prime }d^{3}\mathbf{q}\left( 
\begin{array}{c}
2\widehat{\mathbf{L}}_{q^{\prime }}\cdot \widehat{\mathbf{L}}_{q}\left( 
\begin{array}{c}
b_{\mathbf{q}^{\prime }\uparrow }^{\dagger }b_{\mathbf{q}^{\prime }\uparrow
}D_{\mathbf{q}\uparrow }^{\dagger }D_{\mathbf{q}\uparrow }+b_{\mathbf{q}%
^{\prime }\uparrow }^{\dagger }b_{\mathbf{q}^{\prime }\uparrow }D_{\mathbf{q}%
\downarrow }^{\dagger }D_{\mathbf{q}\downarrow } \\ 
+b_{\mathbf{q}^{\prime }\downarrow }^{\dagger }b_{\mathbf{q}^{\prime
}\downarrow }D_{\mathbf{q}\uparrow }^{\dagger }D_{\mathbf{q}\uparrow }+b_{%
\mathbf{q}^{\prime }\downarrow }^{\dagger }b_{\mathbf{q}^{\prime }\downarrow
}D_{\mathbf{q}\downarrow }^{\dagger }D_{\mathbf{q}\downarrow }
\end{array}
\right) \\ 
+\frac{1}{2}\left( b_{q^{\prime }\uparrow }^{\dagger }b_{q^{\prime }\uparrow
}D_{q\uparrow }^{\dagger }D_{q\uparrow }-b_{q^{\prime }\uparrow }^{\dagger
}b_{q^{\prime }\uparrow }D_{q\downarrow }^{\dagger }D_{q\downarrow }\right)
\\ 
-\frac{1}{2}\left( b_{q^{\prime }\downarrow }^{\dagger }b_{q^{\prime
}\downarrow }D_{q\uparrow }^{\dagger }D_{q\uparrow }-b_{q^{\prime
}\downarrow }^{\dagger }b_{q^{\prime }\downarrow }D_{q\downarrow }^{\dagger
}D_{q\downarrow }\right) \\ 
+b_{q^{\prime }\uparrow }^{\dagger }b_{q^{\prime }\downarrow }D_{q\downarrow
}^{\dagger }D_{q\uparrow }+b_{q^{\prime }\downarrow }^{\dagger }b_{q^{\prime
}\uparrow }D_{q\uparrow }^{\dagger }D_{q\downarrow } \\ 
+\widehat{L}_{q^{\prime }+}\left( 
\begin{array}{c}
b_{\mathbf{q}^{\prime }\uparrow }^{\dagger }b_{\mathbf{q}^{\prime }\uparrow
}D_{\mathbf{q}\downarrow }^{\dagger }D_{\mathbf{q}\uparrow }+b_{\mathbf{q}%
^{\prime }\downarrow }^{\dagger }b_{\mathbf{q}^{\prime }\downarrow }D_{%
\mathbf{q}\downarrow }^{\dagger }D_{\mathbf{q}\uparrow } \\ 
+b_{\mathbf{q}\downarrow }^{\dagger }b_{\mathbf{q}\uparrow }D_{\mathbf{q}%
^{\prime }\uparrow }^{\dagger }D_{\mathbf{q}^{\prime }\uparrow }+b_{\mathbf{q%
}\downarrow }^{\dagger }b_{\mathbf{q}\uparrow }D_{\mathbf{q}^{\prime
}\downarrow }^{\dagger }D_{\mathbf{q}^{\prime }\downarrow }
\end{array}
\right) \\ 
+\widehat{L}_{q^{\prime }-}\left( 
\begin{array}{c}
b_{\mathbf{q}^{\prime }\uparrow }^{\dagger }b_{\mathbf{q}^{\prime }\uparrow
}D_{\mathbf{q}\uparrow }^{\dagger }D_{\mathbf{q}\downarrow }+b_{\mathbf{q}%
^{\prime }\downarrow }^{\dagger }b_{\mathbf{q}^{\prime }\downarrow }D_{%
\mathbf{q}\uparrow }^{\dagger }D_{\mathbf{q}\downarrow } \\ 
+b_{\mathbf{q}\uparrow }^{\dagger }b_{\mathbf{q}\downarrow }D_{\mathbf{q}%
^{\prime }\uparrow }^{\dagger }D_{\mathbf{q}^{\prime }\uparrow }+b_{\mathbf{q%
}\uparrow }^{\dagger }b_{\mathbf{q}\downarrow }D_{\mathbf{q}^{\prime
}\downarrow }^{\dagger }D_{\mathbf{q}^{\prime }\downarrow }
\end{array}
\right) \\ 
+\left( \widehat{L}_{q^{\prime }3}+\widehat{L}_{q3}\right) \left( b_{\mathbf{%
q}^{\prime }\uparrow }^{\dagger }b_{\mathbf{q}^{\prime }\uparrow }D_{\mathbf{%
q}\uparrow }^{\dagger }D_{\mathbf{q}\uparrow }-b_{\mathbf{q}^{\prime
}\downarrow }^{\dagger }b_{\mathbf{q}^{\prime }\downarrow }D_{\mathbf{q}%
\downarrow }^{\dagger }D_{\mathbf{q}\downarrow }\right) \\ 
-\left( \widehat{L}_{q^{\prime }3}-\widehat{L}_{q3}\right) \left( b_{\mathbf{%
q}^{\prime }\uparrow }^{\dagger }b_{\mathbf{q}^{\prime }\uparrow }D_{\mathbf{%
q}\downarrow }^{\dagger }D_{\mathbf{q}\downarrow }-b_{\mathbf{q}^{\prime
}\downarrow }^{\dagger }b_{\mathbf{q}^{\prime }\downarrow }D_{\mathbf{q}%
\uparrow }^{\dagger }D_{\mathbf{q}\uparrow }\right)
\end{array}
\right) ,
\end{eqnarray}
where 
\begin{equation}
\widehat{L}_{q+}=\widehat{L}_{q1}+i\widehat{L}_{q2},\;\;\;\;\;\;\;\widehat{L}%
_{q-}=\widehat{L}_{q1}-i\widehat{L}_{q2}.
\end{equation}
The formulae (137) and (139) apply to the particle-antiparticle system [10]
as well if the operators $D^{\dagger }$ and $D$ are formally replaced by $%
d^{\dagger }$ and $d$.\ 

The requirements (36) with trial state in the form of (24) lead to the
system of equations ($F_{s_{1}s_{2}}\left( \mathbf{p}\right) \equiv
F_{s_{1}s_{2}}$): 
\begin{eqnarray}
\left( \widehat{L}_{3}+1\right) F_{11} &=&m_{J}F_{11}  \notag \\
\widehat{L}_{3}F_{12} &=&m_{J}F_{12}  \notag \\
\widehat{L}_{3}F_{21} &=&m_{J}F_{21} \\
\left( \widehat{L}_{3}-1\right) F_{22} &=&m_{J}F_{22}  \notag
\end{eqnarray}

\begin{eqnarray}
\left( J(J+1)-\widehat{\mathbf{L}}^{2}-2-2\widehat{L}_{3}\right) F_{11} &=&%
\widehat{L}_{-}\left( F_{12}+F_{21}\right)  \notag \\
\left( J(J+1)-\widehat{\mathbf{L}}^{2}-1\right) F_{12} &=&F_{21}+\widehat{L}%
_{+}F_{11}+\widehat{L}_{-}F_{22}  \notag \\
\left( J(J+1)-\widehat{\mathbf{L}}^{2}-1\right) F_{21} &=&F_{12}+\widehat{L}%
_{+}F_{11}+\widehat{L}_{-}F_{22} \\
\left( J(J+1)-\widehat{\mathbf{L}}^{2}-2+2\widehat{L}_{3}\right) F_{22} &=&%
\widehat{L}_{+}\left( F_{12}+F_{21}\right)  \notag
\end{eqnarray}
After substitution of the functions $F_{s_{1}s_{2}}$,\ Eq. (38), into the
system (141) and (142)\ we get 
\begin{equation}
m_{12}=m_{21}=m_{J},\;\ \ \ \ m_{11}=m_{J}-1,\;\ \ \ \ \ \ m_{22}=m_{J}+1,
\end{equation}
\begin{equation}
\ell _{11}=\ell _{22}=\ell _{12}=\ell _{21}\equiv \ell ,
\end{equation}
and 
\begin{eqnarray}
\left( J(J+1)-\ell (\ell +1)-2m_{J}\right) f_{11}^{\ell }(p) &=&\sqrt{(\ell
-m_{J}+1)(\ell +m_{J})}f_{12}^{\ell }(p)  \notag \\
&&+\sqrt{(\ell -m_{J}+1)(\ell +m_{J})}f_{21}^{\ell }(p) \\
\left( J(J+1)-\ell (\ell +1)-1\right) f_{12}^{\ell }(p) &=&f_{21}^{\ell }(p)
\notag \\
&&+\sqrt{(\ell +m_{J})(\ell -m_{J}+1)}f_{11}^{\ell }(p)  \notag \\
&&+\sqrt{(\ell -m_{J})(\ell +m_{J}+1)}f_{22}^{\ell }(p) \\
\left( J(J+1)-\ell (\ell +1)-1\right) f_{21}^{\ell }(p) &=&f_{12}^{\ell }(p)
\notag \\
&&+\sqrt{(\ell +m_{J})(\ell -m_{11}+1)}f_{11}^{\ell }(p)  \notag \\
&&+\sqrt{(\ell -m_{J})(\ell +m_{J}+1)}f_{22}^{\ell }(p) \\
\left( J(J+1)-\ell (\ell +1)+2m_{J}\right) f_{22}^{\ell }(p) &=&\sqrt{(\ell
+m_{J}+1)(\ell -m_{J})}f_{12}^{\ell }(p)  \notag \\
&&+\sqrt{(\ell +m_{J}+1)(\ell -m_{J})}f_{21}^{\ell }(p)
\end{eqnarray}
The solution of this system leads to two categories of relations among the
functions $f_{s_{1}s_{2}}^{\ell }(p)$. The first category, which we call the
trivial one, is obtained when $f_{11}^{\ell }\left( p\right) =f_{22}^{\ell
}\left( p\right) =0$. In this case, as it easy to see from (145) and (148),
we get $f_{12}^{\ell }\left( p\right) =-f_{21}^{\ell }\left( p\right) $.
This solution corresponds to the singlet states of the system with $\ell =J$
($J\geq 0$) (as follows from (146) or (147)). This simple relation allows us
to write the general formula for the components $f_{s_{1}s_{2}}^{\ell
}\left( p\right) $ in the following form

\begin{equation}
f_{s_{1}s_{2}}^{\ell }\left( p\right) =C^{\left( sg\right)
m_{s_{1}s_{2}}}f^{J}\left( p\right) ,
\end{equation}
where the radial function $f^{J}\left( p\right) $ is common for all
components, and the coefficients $C^{\left( sg\right) m_{s_{1}s_{2}}}$ have
the following properties 
\begin{equation}
C^{\left( sg\right) m_{11}}=C^{\left( sg\right) m_{22}}=0,\;\;\;C^{\left(
sg\right) m_{21}}=-C^{\left( sg\right) m_{12}}=1.
\end{equation}
The formulas (149) and (150) with the notation $f_{s_{1}s_{2}}^{\ell }\left(
p\right) \equiv f_{s_{1}s_{2}}^{\left( sg\right) J}\left( p\right) $ make
eq. (41) evident.

The second category corresponds to the triplet states. First of all, it is
not difficult to see from (146) and (147) that $f_{12}^{\ell }\left(
p\right) =f_{21}^{\ell }\left( p\right) \equiv f^{\ell }\left( p\right) $,
and after some simple calculations we get:

for\ $\ell =J-1\;\left( J\geq 1\right) $

\begin{eqnarray}
\left( J-m_{J}\right) f_{11}^{J-1}(p) &=&\sqrt{\left( J-m_{J}\right)
(J+m_{J}-1)}f^{J-1}(p) \\
\left( J+m_{J}\right) f_{22}^{J-1}(p) &=&\sqrt{\left( J+m_{J}\right)
(J-m_{J}-1)}f^{J-1}(p),
\end{eqnarray}
for\ $\ell =J\;\left( J\geq 1\right) $

\begin{eqnarray}
m_{J}f_{11}^{J}(p) &=&-\sqrt{\left( J+m_{J}\right) (J-m_{J}+1)}f^{J}(p) \\
m_{J}f_{22}^{J}(p) &=&\sqrt{\left( J-m_{J}\right) (J+m_{J}+1)}f^{J}(p),
\end{eqnarray}
for\ $\ell =J+1\;\left( J\geq 0\right) $

\begin{eqnarray}
\left( J+1+m_{J}\right) f_{11}^{J+1}(p) &=&-\sqrt{\left( J-m_{J}+2\right)
\left( J+m_{J}+1\right) }f^{J+1}(p) \\
\left( J+1-m_{J}\right) f_{22}^{J+1}(p) &=&-\sqrt{\left( J-m_{J}+1\right)
\left( J+m_{J}+2\right) }f^{J+1}(p).
\end{eqnarray}
It is convenient to introduce the table of coefficients $C_{Jm_{J}}^{\left(
tr\right) \ell m_{s}}$, which represent the relations in (151)-(156).

{\normalsize \vskip0.4truecm }

\begin{center}
Table 5. The C-G coefficients for triplet states (total spin $S=1)$.

\begin{tabular}{||c|c|c|c||}
\hline
& $m_{s}=+1$ & $m_{s}=0$ & $m_{s}=-1$ \\ \hline
$\ell =J-1$ & $\left( \frac{\left( J+m_{J}-1\right) (J+m_{J})}{J\left(
2J-1\right) }\right) ^{1/2}$ & $\left( \frac{\left( J-m_{J}\right) \left(
J+m_{J}\right) }{J\left( 2J-1\right) }\right) ^{1/2}$ & $\left( \frac{\left(
J-m_{J}-1\right) (J-m_{J})}{J\left( 2J-1\right) }\right) ^{1/2}$ \\ \hline
$\ell =J$ & $-\left( \frac{\left( J+m_{J}\right) (J-m_{J}+1)}{J\left(
J+1\right) }\right) ^{1/2}$ & $\frac{m_{J}}{\left( J\left( J+1\right)
\right) ^{1/2}}$ & $\left( \frac{\left( J-m_{J}\right) (J+m_{J}+1)}{J\left(
J+1\right) }\right) ^{1/2}$ \\ \hline
$\ell =J+1$ & $\left( \frac{\left( J-m_{J}+1\right) \left( J-m_{J}+2\right) 
}{\left( J+1\right) \left( 2J+3\right) }\right) ^{1/2}$ & $-\left( \frac{%
\left( J-m_{J}+1\right) \left( J+m_{J}+1\right) }{\left( J+1\right) \left(
2J+3\right) }\right) ^{1/2}$ & $\left( \frac{\left( J+m_{J}+2\right) \left(
J+m_{J}+1\right) }{\left( J+1\right) \left( 2J+3\right) }\right) ^{1/2}$ \\ 
\hline
\end{tabular}
\end{center}

Thus, we can write the relations between the components $f_{s_{1}s_{2}}^{%
\ell }\left( p\right) $ in the compact form 
\begin{equation}
f_{s_{1}s_{2}}^{\ell }\left( p\right) =C_{Jm_{J}}^{\left( tr\right) \ell
m_{s}}f^{\ell }\left( p\right) .
\end{equation}
The coefficients $C_{Jm_{J}}^{\left( tr\right) \ell m_{s}}$\ coincide with
the usual Clebsch-Gordan coefficients for total spin $S=1$ except for a
factor $2$ in the denominator. The expression (38) can now be written in an
explicit form

\begin{equation}
F_{s_{1}s_{2}}(\mathbf{p})=C_{Jm_{J}}^{\left( tr\right)
(J-1)m_{s}}f^{J-1}(p)Y_{J-1}^{m_{s_{1}s_{2}}}(\widehat{\mathbf{p}}%
)+C_{Jm_{J}}^{\left( tr\right) (J)m_{s}}f^{J}(p)Y_{J}^{m_{s_{1}s_{2}}}(%
\widehat{\mathbf{p}})+C_{Jm_{J}}^{\left( tr\right)
(J+1)m_{s}}f^{J+1}(p)Y_{J+1}^{m_{s_{1}s_{2}}}(\widehat{\mathbf{p}}).
\end{equation}

However, as is shown in Appendix B, the first and third terms have parity $%
P=\left( -1\right) ^{J}$, while the second term has parity $P=\left(
-1\right) ^{J+1}$. Thus, we get the result (56) by suppressing the second
term in (158). The second term in (158) is associated with the singlet
solution (42) for the mixed-spin states, which have the same parity.

\vskip0.8truecm {\normalsize \noindent {\textbf{\large Appendix B: Parity
and charge conjugation}} }

{\normalsize \vskip0.4truecm }

{\normalsize 
}

We consider the application of the parity operator to the trial state (24): 
\begin{equation}
\widehat{\mathcal{P}}\,\left| \psi _{trial}\right\rangle =\underset{%
s_{1}s_{2}}{\sum }\int d^{3}\mathbf{p}F_{s_{1}s_{2}}(\mathbf{p})\widehat{%
\mathcal{P}}b_{ps_{1}}^{\dagger }D_{-ps_{2}}^{\dagger }\,\left|
0\right\rangle =\underset{s_{1}s_{2}}{\sum }\int d^{3}\mathbf{p}%
F_{s_{1}s_{2}}(\mathbf{p})\widehat{\mathcal{P}}b_{ps_{1}}^{\dagger }\widehat{%
\mathcal{P}}^{-1}\widehat{\mathcal{P}}D_{-ps_{2}}^{\dagger }\widehat{%
\mathcal{P}}^{-1}\widehat{\mathcal{P}}\,\left| 0\right\rangle .
\end{equation}
Making use of the properties 
\begin{equation}
\widehat{\mathcal{P}}b_{ps_{1}}^{\dagger }\widehat{\mathcal{P}}^{-1}=\eta
^{P}b_{-ps_{1}}^{\dagger },\;\;\;\;\;\widehat{\mathcal{P}}%
D_{-ps_{2}}^{\dagger }\widehat{\mathcal{P}}^{-1}=-\eta
^{P}D_{ps_{2}}^{\dagger },\;\;\;\;\;\widehat{\mathcal{P}}\,\left|
0\right\rangle =\,\left| 0\right\rangle ,
\end{equation}
where $\eta ^{P}$ is the intrinsic parity ($\left( \eta ^{P}\right) ^{2}=1$%
), it follows that 
\begin{equation*}
\widehat{\mathcal{P}}\,\left| \psi _{trial}\right\rangle =\underset{%
s_{1}s_{2}}{\sum }\int d^{3}\mathbf{p}F_{s_{1}s_{2}}(\mathbf{p})\widehat{%
\mathcal{P}}b_{ps_{1}}^{\dagger }D_{-ps_{2}}^{\dagger }\,\left|
0\right\rangle =\underset{s_{1}s_{2}}{-\sum }\int d^{3}\mathbf{p}%
F_{s_{1}s_{2}}(-\mathbf{p})b_{ps_{1}}^{\dagger }D_{-ps_{2}}^{\dagger
}\,\left| 0\right\rangle
\end{equation*}
\begin{equation}
=P\underset{s_{1}s_{2}}{\sum }\int d^{3}\mathbf{p}F_{s_{1}s_{2}}(\mathbf{p}%
)b_{ps_{1}}^{\dagger }D_{-ps_{2}}^{\dagger }\,\left| 0\right\rangle ,
\end{equation}
where the parity eigenvalue $P$ depends on the symmetry of $F_{s_{1}s_{2}}(%
\mathbf{p})$ in different states. For the singlet states\textbf{\ }$\left(
\ell =J\right) $ we get from (36) $F_{s_{1}s_{2}}(-\mathbf{p})=\left(
-1\right) ^{J}F_{s_{1}s_{2}}(\mathbf{p})$, so $P=\left( -1\right) ^{J+1}$.
For the triplet states with\textbf{\ }$\ell =J$ we get from (47) $%
F_{s_{1}s_{2}}(-\mathbf{p})=\left( -1\right) ^{J}F_{s_{1}s_{2}}(\mathbf{p})$%
, so $P=\left( -1\right) ^{J+1}$. For the triplet states with\textbf{\ }$%
\ell =J\pm 1$ we get from (56) $F_{s_{1}s_{2}}(-\mathbf{p})=\left( -1\right)
^{J+1}F_{s_{1}s_{2}}(\mathbf{p})$, so $P=\left( -1\right) ^{J}$.

Charge conjugation is associated with the interchange of the particle and
antiparticle. Applying the charge conjugation operator to the trial state of
the form (24) we get 
\begin{equation}
\widehat{\mathcal{C}}\,\left| \psi _{trial}\right\rangle =\underset{%
s_{1}s_{2}}{\sum }\int d^{3}\mathbf{p}F_{s_{1}s_{2}}(\mathbf{p})\widehat{%
\mathcal{C}}b_{ps_{1}}^{\dagger }d_{-ps_{2}}^{\dagger }\,\left|
0\right\rangle =\underset{s_{1}s_{2}}{\sum }\int d^{3}\mathbf{p}%
F_{s_{1}s_{2}}(\mathbf{p})\widehat{\mathcal{C}}b_{ps_{1}}^{\dagger }\widehat{%
\mathcal{C}}^{-1}\widehat{\mathcal{C}}d_{-ps_{2}}^{\dagger }\widehat{%
\mathcal{C}}^{-1}\widehat{\mathcal{C}}\,\left| 0\right\rangle .
\end{equation}
Using the relations 
\begin{equation}
\widehat{\mathcal{C}}b_{ps_{1}}^{\dagger }\widehat{\mathcal{C}}^{-1}=\eta
^{C}d_{ps_{1}}^{\dagger },\;\;\;\;\;\widehat{\mathcal{C}}d_{-ps_{2}}^{%
\dagger }\widehat{\mathcal{C}}^{-1}=\eta ^{C}b_{-ps_{2}}^{\dagger
},\;\;\;\;\;\widehat{\mathcal{C}}\,\left| 0\right\rangle =\,\left|
0\right\rangle ,
\end{equation}
where $\left( \eta ^{C}\right) ^{2}=1$, we obtain 
\begin{equation*}
\widehat{\mathcal{C}}\,\left| \psi _{trial}\right\rangle =\underset{%
s_{1}s_{2}}{\sum }\int d^{3}\mathbf{p}F_{s_{1}s_{2}}(\mathbf{p})\widehat{%
\mathcal{C}}b_{ps_{1}}^{\dagger }d_{-ps_{2}}^{\dagger }\,\left|
0\right\rangle =-\underset{s_{1}s_{2}}{\sum }\int d^{3}\mathbf{p}%
F_{s_{2}s_{1}}(\mathbf{p})b_{ps_{1}}^{\dagger }d_{-ps_{2}}^{\dagger
}\,\left| 0\right\rangle
\end{equation*}
\begin{equation}
=C\underset{s_{1}s_{2}}{\sum }\int d^{3}\mathbf{p}F_{s_{1}s_{2}}(\mathbf{p}%
)b_{ps_{1}}^{\dagger }d_{-ps_{2}}^{\dagger }\,\left| 0\right\rangle ,
\end{equation}
where the charge conjugation quantum number $C$ depends on the symmetry of $%
F_{s_{1}s_{2}}(\mathbf{p})$ in different states.

For the singlet states\textbf{\ }$\left( \ell =J\right) $ we get from (42) $%
F_{s_{1}s_{2}}(-\mathbf{p})=\left( -1\right) ^{J+1}F_{s_{1}s_{2}}(\mathbf{p}%
) $, so $C=\left( -1\right) ^{J}$.

For the triplet states with\textbf{\ }$\ell =J$ we get from (47) $%
F_{s_{1}s_{2}}(-\mathbf{p})=\left( -1\right) ^{J}F_{s_{1}s_{2}}(\mathbf{p})$%
, so $C=\left( -1\right) ^{J+1}$.

For the triplet states with\textbf{\ }$\ell =J\pm 1$ we get from (56) $%
F_{s_{1}s_{2}}(-\mathbf{p})=\left( -1\right) ^{J+1}F_{s_{1}s_{2}}(\mathbf{p}%
) $, so $C=\left( -1\right) ^{J}$.

\vskip0.8truecm {\normalsize \noindent {\textbf{{\large Appendix C:
Expansion of the spinors and $\mathcal{M}$-matrix elements}}} }

{\normalsize \vskip0.4truecm }

{\normalsize 
}

We recall the form of the particle spinors: 
\begin{equation}
u(\mathbf{p,}i)=N_{\mathbf{p}}\left[ 
\begin{array}{c}
1 \\ 
\frac{(\overrightarrow{\sigma }\mathbf{\cdot p})}{\omega _{p}+m_{1}}
\end{array}
\right] \varphi _{i},
\end{equation}
where 
\begin{equation}
\varphi _{1}=\left[ 
\begin{array}{c}
1 \\ 
0
\end{array}
\right] ,\;\;\;\varphi _{2}=\left[ 
\begin{array}{c}
0 \\ 
1
\end{array}
\right] ,\;\;\;\;\;N_{\mathbf{p}}=\sqrt{\frac{\omega _{p}+m_{1}}{2m_{1}}}.
\end{equation}
The antiparticle or ``positron'' representation for the $v_{i}(\mathbf{p})$
spinors has the form

\begin{equation}
v(\mathbf{p,}i)=N_{\mathbf{p}}\left[ 
\begin{array}{c}
\frac{(\overrightarrow{\sigma }\mathbf{\cdot p})}{\omega _{p}+m_{1}} \\ 
1
\end{array}
\right] \chi _{i},
\end{equation}
where 
\begin{equation}
\chi _{1}=\left[ 
\begin{array}{c}
0 \\ 
1
\end{array}
\right] ,\;\;\;\;\;\chi _{2}=-\left[ 
\begin{array}{c}
1 \\ 
0
\end{array}
\right] .
\end{equation}
The normalization is 
\begin{equation}
\overline{u}(\mathbf{p,}i)u(\mathbf{p,}j)=\delta _{ij},\;\;\;\;\;\;\;%
\overline{v}(\mathbf{p,}i)v(\mathbf{p,}j)=-\delta _{ij}.
\end{equation}
Expanding in powers of $p/m_{1}$ and keeping the lowest order non-trivial
terms,

\begin{equation}
\frac{(\overrightarrow{\sigma }\mathbf{\cdot p})}{\omega _{p}+m_{1}}\simeq 
\frac{(\overrightarrow{\sigma }\mathbf{\cdot p})}{2m_{1}},
\end{equation}
\begin{equation}
N_{\mathbf{p}}=\sqrt{\frac{\omega _{p}+m_{1}}{2m_{1}}}\simeq 1+\frac{\mathbf{%
p}^{2}}{8m_{1}^{2}},
\end{equation}
we obtain the result 
\begin{equation}
u(\mathbf{p,}i)\simeq \left( 1+\frac{\mathbf{p}^{2}}{8m_{1}^{2}}\right) %
\left[ 
\begin{array}{c}
1 \\ 
\frac{(\overrightarrow{\sigma }\mathbf{\cdot p})}{2m_{1}}
\end{array}
\right] \varphi _{i}=\left[ 
\begin{array}{c}
\left( 1+\frac{\mathbf{p}^{2}}{8m_{1}^{2}}\right) \\ 
\frac{(\overrightarrow{\sigma }\mathbf{\cdot p})}{2m_{1}}
\end{array}
\right] \varphi _{i},
\end{equation}
\begin{equation}
v(\mathbf{p,}i)\simeq \left( 1+\frac{\mathbf{p}^{2}}{8m_{1}^{2}}\right) 
\left[ 
\begin{array}{c}
\frac{(\overrightarrow{\sigma }\mathbf{\cdot p})}{2m_{1}} \\ 
1
\end{array}
\right] \chi _{i}=\left[ 
\begin{array}{c}
\frac{(\overrightarrow{\sigma }\mathbf{\cdot p})}{2m_{1}} \\ 
\left( 1+\frac{\mathbf{p}^{2}}{8m_{1}^{2}}\right)
\end{array}
\right] \chi _{i}.
\end{equation}
The $\mathcal{M}$- matrix elements (69)-(72) have the following nonzero
components:\ 

Orbital 
\begin{eqnarray}
\mathcal{M}_{1111}^{\left( orb\right) }(\mathbf{q},\mathbf{p}) &=&\mathcal{M}%
_{1212}^{\left( orb\right) }(\mathbf{q},\mathbf{p})=\mathcal{M}%
_{2121}^{\left( orb\right) }(\mathbf{q},\mathbf{p})=\mathcal{M}%
_{2222}^{\left( orb\right) }(\mathbf{q},\mathbf{p}) \\
&=&iq_{1}q_{2}\left( 
\begin{array}{c}
\frac{1}{\left( \mathbf{p-q}\right) ^{2}}+\frac{1}{8m_{1}m_{2}}\left( \frac{%
m_{2}}{m_{1}}+\frac{m_{1}}{m_{2}}\right) -\frac{\left( \left( \mathbf{p-q}%
\right) \mathbf{\cdot p}\right) ^{2}}{m_{1}m_{2}\left( \mathbf{p-q}\right)
^{4}} \\ 
+\frac{\mathbf{q}^{2}}{m_{1}m_{2}\left( \mathbf{p-q}\right) ^{2}}+\left( 
\frac{m_{2}}{m_{1}}+\frac{m_{1}}{m_{2}}\right) \frac{\mathbf{q\cdot p}}{%
2m_{1}m_{2}\left( \mathbf{p-q}\right) ^{2}}
\end{array}
\right) ,  \notag
\end{eqnarray}

Spin-orbit 
\begin{equation}
\mathcal{M}_{1111}^{\left( s-o\right) }(\mathbf{q},\mathbf{p})=\frac{%
iq_{1}q_{2}}{8m_{1}m_{2}}\left( \frac{m_{2}}{m_{1}}+\frac{m_{1}}{m_{2}}%
+4\right) \frac{p_{-}q_{+}-p_{+}q_{-}}{\left( \mathbf{p-q}\right) ^{2}},
\end{equation}
\begin{equation}
\mathcal{M}_{2222}^{\left( s-o\right) }(\mathbf{q},\mathbf{p})=\frac{%
iq_{1}q_{2}}{8m_{1}m_{2}}\left( \frac{m_{2}}{m_{1}}+\frac{m_{1}}{m_{2}}%
+4\right) \frac{p_{+}q_{-}-p_{-}q_{+}}{\left( \mathbf{p-q}\right) ^{2}},
\end{equation}
\begin{equation}
\mathcal{M}_{1112}^{\left( s-o\right) }(\mathbf{q},\mathbf{p})=\mathcal{M}%
_{2122}^{\left( s-o\right) }(\mathbf{q},\mathbf{p})=\frac{iq_{1}q_{2}}{%
4m_{1}m_{2}}\left( \frac{m_{1}}{m_{2}}+2\right) \frac{p_{3}q_{-}-p_{-}q_{3}}{%
\left( \mathbf{p-q}\right) ^{2}},
\end{equation}
\begin{equation}
\mathcal{M}_{1211}^{\left( s-o\right) }(\mathbf{q},\mathbf{p})=\mathcal{M}%
_{2221}^{\left( s-o\right) }(\mathbf{q},\mathbf{p})=\frac{iq_{1}q_{2}}{%
4m_{1}m_{2}}\left( \frac{m_{1}}{m_{2}}+2\right) \frac{p_{+}q_{3}-p_{3}q_{+}}{%
\left( \mathbf{p-q}\right) ^{2}},
\end{equation}
\begin{equation}
\mathcal{M}_{1121}^{\left( s-o\right) }(\mathbf{q},\mathbf{p})=\mathcal{M}%
_{1222}^{\left( s-o\right) }(\mathbf{q},\mathbf{p})=\frac{iq_{1}q_{2}}{%
4m_{1}m_{2}}\left( \frac{m_{2}}{m_{1}}+2\right) \frac{p_{3}q_{-}-p_{-}q_{3}}{%
\left( \mathbf{p-q}\right) ^{2}},
\end{equation}
\begin{equation}
\mathcal{M}_{2111}^{\left( s-o\right) }(\mathbf{q},\mathbf{p})=\mathcal{M}%
_{2212}^{\left( s-o\right) }(\mathbf{q},\mathbf{p})=\frac{iq_{1}q_{2}}{%
4m_{1}m_{2}}\left( \frac{m_{2}}{m_{1}}+2\right) \frac{p_{+}q_{3}-p_{3}q_{+}}{%
\left( \mathbf{p-q}\right) ^{2}},
\end{equation}
\begin{equation}
\mathcal{M}_{1212}^{\left( s-o\right) }(\mathbf{q},\mathbf{p})=-\mathcal{M}%
_{2121}^{\left( s-o\right) }(\mathbf{q},\mathbf{p})=\frac{iq_{1}q_{2}}{%
8m_{1}m_{2}}\left( \frac{m_{2}}{m_{1}}-\frac{m_{1}}{m_{2}}\right) \frac{%
p_{-}q_{+}-p_{+}q_{-}}{\left( \mathbf{p-q}\right) ^{2}},
\end{equation}

Spin-spin 
\begin{eqnarray}
\mathcal{M}_{1111}^{\left( s-s\right) }(\mathbf{q},\mathbf{p}) &=&\mathcal{M}%
_{2222}^{\left( s-s\right) }(\mathbf{q},\mathbf{p})=-\mathcal{M}%
_{1212}^{\left( s-s\right) }(\mathbf{q},\mathbf{p})=\left( \mathcal{M}%
_{2121}^{\left( s-s\right) }(\mathbf{q},\mathbf{p})\right) ^{\ast } \\
&=&\frac{iq_{1}q_{2}}{4m_{1}m_{2}}\left( \frac{\left( q_{3}\mathbf{-}%
p_{3}\right) ^{2}}{\left( \mathbf{q-p}\right) ^{2}}-1\right) ,  \notag
\end{eqnarray}
\begin{eqnarray}
\mathcal{M}_{1112}^{\left( s-s\right) }(\mathbf{q},\mathbf{p}) &=&\left( 
\mathcal{M}_{2221}^{\left( s-s\right) }(\mathbf{q},\mathbf{p})\right) ^{\ast
}=-\left( \mathcal{M}_{1211}^{\left( s-s\right) }(\mathbf{q},\mathbf{p}%
)\right) ^{\ast }=-\mathcal{M}_{2122}^{\left( s-s\right) }(\mathbf{q},%
\mathbf{p})  \notag \\
\mathcal{M}_{1121}^{\left( s-s\right) }(\mathbf{q},\mathbf{p}) &=&\left( 
\mathcal{M}_{2212}^{\left( s-s\right) }(\mathbf{q},\mathbf{p})\right) ^{\ast
}=-\mathcal{M}_{1222}^{\left( s-s\right) }(\mathbf{q},\mathbf{p})=-\left( 
\mathcal{M}_{2111}^{\left( s-s\right) }(\mathbf{q},\mathbf{p})\right) ^{\ast
} \\
&=&\frac{iq_{1}q_{2}}{4m_{1}m_{2}}\left( \frac{%
q_{-}q_{3}-p_{-}q_{3}-p_{3}q_{-}+p_{-}p_{3}}{\left( \mathbf{q-p}\right) ^{2}}%
\right) ,  \notag
\end{eqnarray}
\begin{equation}
\mathcal{M}_{1122}^{\left( s-s\right) }(\mathbf{q},\mathbf{p})=-\left( 
\mathcal{M}_{2211}^{\left( s-s\right) }(\mathbf{q},\mathbf{p})\right) ^{\ast
}=\frac{iq_{1}q_{2}}{4m_{1}m_{2}}\frac{q_{-}^{2}-2q_{-}p_{-}+p_{-}^{2}}{%
\left( \mathbf{q-p}\right) ^{2}},
\end{equation}
\begin{equation}
\mathcal{M}_{1221}^{\left( s-s\right) }(\mathbf{q},\mathbf{p})=\mathcal{M}%
_{2112}^{\left( s-s\right) }(\mathbf{q},\mathbf{p})=\frac{iq_{1}q_{2}}{%
4m_{1}m_{2}}\left( \frac{q_{+}q_{-}-p_{+}q_{-}-p_{-}q_{+}+p_{+}p_{-}}{\left( 
\mathbf{q-p}\right) ^{2}}-2\right) .
\end{equation}
Here $\mathbf{p=}\left( p_{1},p_{2},p_{3}\right) $ and 
\begin{eqnarray}
p_{+} &=&p_{1}+ip_{2}=-\sqrt{\frac{8\pi }{3}}pY_{1}^{1}\left( \theta
,\varphi \right) ,  \notag \\
p_{-} &=&p_{1}-ip_{2}=\sqrt{\frac{8\pi }{3}}pY_{1}^{-1}\left( \theta
,\varphi \right) , \\
p_{3} &=&\sqrt{\frac{4\pi }{3}}pY_{1}^{0}\left( \theta ,\varphi \right) .
\end{eqnarray}

For a particle-antiparticle system the annihilation components are 
\begin{equation}
\mathcal{M}_{1111}^{anh}=\mathcal{M}_{2222}^{anh}=\frac{ie^{2}}{2m^{2}},
\end{equation}
\begin{equation}
\mathcal{M}_{1212}^{anh}=\mathcal{M}_{1221}^{anh}=\mathcal{M}_{2112}^{anh}=%
\mathcal{M}_{2121}^{anh}=\frac{ie^{2}}{4m^{2}}.
\end{equation}
\vskip0.8truecm {\normalsize \noindent {\textbf{\large Appendix D: Some
useful expressions, identities and integrals}} }

{\normalsize \vskip0.4truecm }

{\normalsize 
}

The following expressions and identities are useful for evaluating the $%
\mathcal{M}$ -matrix: 
\begin{equation}
\frac{1}{(\mathbf{q-p})^{2}}=\frac{2\pi }{\left| \mathbf{p}\right| \left| 
\mathbf{q}\right| }\sum_{\lambda }Q_{\lambda }(z)\sum_{m_{\lambda }=-\lambda
}^{+\lambda }Y_{\lambda }^{m_{\lambda }}(\widehat{\mathbf{p}})Y_{\lambda
}^{m_{\lambda }\ast }(\widehat{\mathbf{q}}),
\end{equation}
where $z=\left( p^{2}+q^{2}\right) /2pq$, and $Q_{\lambda }(z)$ is the
Legendre function of the second kind of order $\lambda $ [17]. Then 
\begin{equation}
\frac{\left( \left( \mathbf{p}-\mathbf{q}\right) \cdot \mathbf{p}\right) ^{2}%
}{\left( \mathbf{p}-\mathbf{q}\right) ^{4}}=\frac{\mathbf{p}^{2}}{\left( 
\mathbf{p}-\mathbf{q}\right) ^{2}}-\frac{\left( \mathbf{p}\times \mathbf{q}%
\right) ^{2}}{\left( \mathbf{p}-\mathbf{q}\right) ^{4}}.
\end{equation}
The angular integration in (58), (59), (64) involves the following integrals 
\begin{equation}
\int d\widehat{\mathbf{p}}\,d\widehat{\mathbf{q}}\,\digamma \left( \widehat{%
\mathbf{p}}\cdot \widehat{\mathbf{q}}\right) Y_{J^{\prime }}^{m_{J}^{\prime
}}(\widehat{\mathbf{q}})Y_{J}^{m_{J}\ast }(\widehat{\mathbf{p}})=2\pi \delta
_{J^{\prime }J}\delta _{m_{J}^{\prime }m_{J}}\int d\left( \widehat{\mathbf{p}%
}\cdot \widehat{\mathbf{q}}\right) \digamma \left( \widehat{\mathbf{p}}\cdot 
\widehat{\mathbf{q}}\right) P_{J}\left( \widehat{\mathbf{p}}\cdot \widehat{%
\mathbf{q}}\right) ,
\end{equation}
\begin{equation}
\int d\left( \widehat{\mathbf{p}}\cdot \widehat{\mathbf{q}}\right) \frac{%
\widehat{\mathbf{p}}\cdot \widehat{\mathbf{q}}}{\left( \mathbf{p}-\mathbf{q}%
\right) ^{2}}P_{J}\left( \widehat{\mathbf{p}}\cdot \widehat{\mathbf{q}}%
\right) =\frac{1}{\left| \mathbf{p}\right| \left| \mathbf{q}\right| }\left( 
\frac{J+1}{2J+1}Q_{J+1}\left( z\right) +\frac{J}{2J+1}Q_{J-1}\left( z\right)
\right) ,
\end{equation}
\begin{equation}
\int d\left( \widehat{\mathbf{p}}\cdot \widehat{\mathbf{q}}\right) \frac{%
\left( \mathbf{p}\times \mathbf{q}\right) ^{2}}{\left( \mathbf{p}-\mathbf{q}%
\right) ^{4}}P_{J}\left( \widehat{\mathbf{p}}\cdot \widehat{\mathbf{q}}%
\right) =\frac{\left( J+1\right) \left( J+2\right) }{2\left( 2J+1\right) }%
Q_{J+1}\left( z\right) -\frac{J\left( J-1\right) }{2\left( 2J+1\right) }%
Q_{J-1}\left( z\right) .
\end{equation}
Here $\digamma \left( \widehat{\mathbf{p}}\cdot \widehat{\mathbf{q}}\right) $
is an arbitrary function of $\widehat{\mathbf{p}}\cdot \widehat{\mathbf{q}}$%
, $P_{J}\left( x\right) $ is the Legendre polynomial of order $J$.

The integrals in the form 
\begin{equation}
\int d\widehat{\mathbf{p}}\,\,Y_{J}^{m_{J}\ast }(\widehat{\mathbf{p}}%
)Y_{J^{\prime }}^{m_{J}^{\prime }}(\widehat{\mathbf{p}})Y_{J^{\prime \prime
}}^{m_{J}^{\prime \prime }}(\widehat{\mathbf{p}})
\end{equation}
can be calculated using the Wigner-Eckart theorem [17].

The calculation of the relativistic energy corrections involves the
integrals 
\begin{equation}
\int_{0}^{\infty }\int_{0}^{\infty }dp\,dq\,p^{2}q^{2}f^{J}(p)f^{J}(q)=2\pi
\left( \frac{\alpha m_{r}}{n}\right) ^{3}\delta _{J,0},
\end{equation}
\begin{equation}
\int_{0}^{\infty }\int_{0}^{\infty }dp\,dq\,pqf^{J}(p)f^{J}(q)Q_{J}(z_{1})=%
\frac{\pi \alpha m_{r}}{n^{2}},
\end{equation}
\begin{eqnarray}
\int_{0}^{\infty }\int_{0}^{\infty
}dp\,dq\,p^{2}q^{2}f^{J}(p)f^{J}(q)Q_{J}(z_{1}) &=&  \notag \\
\int_{0}^{\infty }\int_{0}^{\infty
}dp\,dq\,p^{3}qf^{J}(p)f^{J}(q)Q_{J}(z_{1}) &=&\pi \left( \frac{\alpha m_{r}%
}{n}\right) ^{3}\left( \frac{4}{2J+1}-\frac{1}{n}\right) ,
\end{eqnarray}
\begin{equation}
\int_{0}^{\infty }\int_{0}^{\infty
}dp\,dq\,p^{2}q^{2}f^{J}(p)f^{J}(q)Q_{J-1}(z_{1})=\pi \left( \frac{\alpha
m_{r}}{n}\right) ^{3}\left( \frac{2}{J}-\frac{1}{n}\right) ,
\end{equation}
\begin{equation}
\int_{0}^{\infty }\int_{0}^{\infty
}dp\,dq\,p^{2}q^{2}f^{J}(p)f^{J}(q)Q_{J+1}\left( z_{1}\right) =\pi \left( 
\frac{\alpha m_{r}}{n}\right) ^{3}\left( \frac{2}{J+1}-\frac{1}{n}\right) .
\end{equation}
Here $f^{J}$ is the nonrelativistic hydrogen-like\ radial wave function in
momentum space [18] 
\begin{equation}
f^{J}(p)\equiv f_{n}^{J}(p)=\left( \frac{2}{\pi }\frac{\left( n-J-1\right) !%
}{\left( n+J\right) !}\right) ^{1/2}\frac{n^{J+2}p^{J}2^{2\left( J+1\right)
}J!}{\left( n^{2}p^{2}+1\right) ^{J+2}}\mathcal{G}_{n-J-1}^{J+1}\left( \frac{%
n^{2}p^{2}-1}{n^{2}p^{2}+1}\right) ,
\end{equation}
where $\mathcal{G}_{n-J-1}^{J+1}\left( x\right) $ are the Gegenbauer
functions.

\vskip0.8truecm {\normalsize \noindent {\textbf{\large Appendix E:
Diagonalization of the expectation value of the Hamiltonian}} }

{\normalsize \vskip0.4truecm }

{\normalsize 
}

The matrix representation of the perturbing Hamiltonian $\Delta \widehat{H}=%
\widehat{H}-\widehat{H}_{NR}-M$, in the basis of the states $\mid sg\rangle $%
, $\mid tr\rangle $\ (48), (49), is ($J\neq 0$) 
\begin{equation}
\left\langle \psi \right| \,\Delta \widehat{H}\,\left| \psi \right\rangle
_{t=0}=\left[ a_{ij}\right] \;\;\;\;(i,j=1,2),
\end{equation}
where the matrix elements $a_{ij}$ are 
\begin{equation}
a_{11}=\left\langle sg\right| \,\Delta \widehat{H}\,\left| sg\right\rangle =-%
\frac{\alpha ^{4}m_{r}}{n^{3}}\left( \frac{1}{2J+1}-\left( 3-\frac{m_{r}}{M}%
\right) \frac{1}{8n}\right) ,
\end{equation}
\begin{equation}
a_{22}=\left\langle tr\right| \,\Delta \widehat{H}\,\left| tr\right\rangle =-%
\frac{\alpha ^{4}m_{r}}{n^{3}}\left( \frac{1}{2J+1}-\frac{1}{2J\left(
J+1\right) \left( 2J+1\right) }-\left( 3-\frac{m_{r}}{M}\right) \frac{1}{8n}%
\right) ,
\end{equation}
\begin{equation}
a_{12}=a_{21}=\left\langle sg\right| \,\Delta \widehat{H}\,\left|
tr\right\rangle =\left\langle tr\right| \,\Delta \widehat{H}\,\left|
sg\right\rangle =\frac{\alpha ^{4}m_{r}}{n^{3}}\frac{m_{1}-m_{2}}{2M}\frac{1%
}{2J+1}\frac{1}{\sqrt{J\left( J+1\right) }}.
\end{equation}
Note that, in the case of positronium, the elements $a_{11}$ and $a_{22}$
give the energy corrections for pure singlet and triplet states
respectively. Diagonalization of this matrix leads to (103) with
eigenvectors (54).

\vskip0.8truecm {\normalsize \noindent {\textbf{{\large Appendix F: $%
\mathcal{K}_{12}$, $\mathcal{K}_{21}$ kernels for \ }}}}$l${\normalsize {%
\textbf{\large $=$}}}$J\mp 1${\normalsize {\textbf{\large \ states}} }

{\normalsize \vskip0.4truecm }

{\normalsize 
}

The contribution to the energy correction due to the kernel $\mathcal{K}%
_{12} $ is 
\begin{equation}
\int dp\,dq\,p^{2}q^{2}\mathcal{K}_{12}\left( p,q\right)
f^{J-1}(p)f^{J+1}(q),
\end{equation}
where

\begin{equation}
\mathcal{K}_{12}\left( p,q\right) =\underset{\sigma _{1}\sigma _{2}s_{1}s_{2}%
}{\sum }C_{Jm_{J}12}^{s_{1}s_{2}\sigma _{1}\sigma _{2}}\int d\widehat{%
\mathbf{p}}\,d\widehat{\mathbf{q}}\,\mathcal{M}_{s_{1}s_{2}\sigma _{1}\sigma
_{2}}^{ope\left( s-s\right) }\left( \mathbf{p,q}\right) Y_{J+1}^{m_{\sigma
_{1}\sigma _{2}}}(\widehat{\mathbf{q}})Y_{J-1}^{m_{s_{1}s_{2}}\ast }(%
\widehat{\mathbf{p}}).
\end{equation}
This requires the following integral 
\begin{equation}
\underset{\sigma _{1}\sigma _{2}s_{1}s_{2}}{\sum }C_{Jm_{J}12}^{s_{1}s_{2}%
\sigma _{1}\sigma _{2}}\int d^{3}\mathbf{p\,}d^{3}\mathbf{q\,}%
f^{J-1}(p)Y_{J-1}^{m_{s_{1}s_{2}}\ast }(\widehat{\mathbf{p}})\mathcal{M}%
_{s_{1}s_{2}\sigma _{1}\sigma _{2}}^{ope\left( s-s\right) }\left( \mathbf{p,q%
}\right) f^{J+1}(q)Y_{J+1}^{m_{\sigma _{1}\sigma _{2}}}(\widehat{\mathbf{q}}%
).
\end{equation}
We calculate this form in coordinate space. The Fourier transforming of $%
\mathcal{M}_{s_{1}s_{2}\sigma _{1}\sigma _{2}}\left( \mathbf{p,q}\right) \;$%
is

\begin{equation}
\mathcal{M}_{s_{1}s_{2}\sigma _{1}\sigma _{2}}\left( \mathbf{p,q}\right)
=\int d^{3}\mathbf{r}d^{3}\mathbf{r}^{\prime }\mathcal{M}_{s_{1}s_{2}\sigma
_{1}\sigma _{2}}\left( \mathbf{r,r}^{\prime }\right) e^{-i\left( \mathbf{p-q}%
\right) \cdot \left( \mathbf{r-r}^{\prime }\right) },
\end{equation}
where the $\mathcal{M}_{s_{1}s_{2}\sigma _{1}\sigma _{2}}\left( \mathbf{r,r}%
^{\prime }\right) \mathcal{\ }$matrix in general is a local operator [16] 
\begin{equation}
\mathcal{M}_{s_{1}s_{2}\sigma _{1}\sigma _{2}}\left( \mathbf{r,r}^{\prime
}\right) =\mathcal{M}_{s_{1}s_{2}\sigma _{1}\sigma _{2}}\left( \mathbf{r}%
\right) \delta \left( \mathbf{r-r}^{\prime }\right) .
\end{equation}
We apply this transformation to the $\mathcal{M}_{s_{1}s_{2}\sigma
_{1}\sigma _{2}}^{ope\left( s-s\right) }\left( \mathbf{p,q}\right) $-matrix
(see (71)). Because of the angular integration in (207), only the first term
of (71) contributes. The\ Fourier transformation of that term is 
\begin{equation}
\frac{\left( \overrightarrow{\sigma }_{2}\cdot \left( \mathbf{p-q}\right)
\right) \left( \overrightarrow{\sigma }_{1}\cdot \left( \mathbf{p-q}\right)
\right) }{4\left( \mathbf{p-q}\right) ^{2}}\;\;\;\rightarrow \;\;\;3\frac{%
\left( \overrightarrow{\sigma }_{2}\cdot \mathbf{r}\right) \left( 
\overrightarrow{\sigma }_{1}\cdot \mathbf{r}\right) }{16\pi r^{5}}.
\end{equation}
Furthermore,

\begin{equation}
\int d^{3}\mathbf{p}f^{J-1}(p)Y_{J-1}^{m_{s_{1}s_{2}}\ast }(\widehat{\mathbf{%
p}})e^{-i\mathbf{p}\cdot \mathbf{r}}=R_{n}^{J-1}(r)Y_{J-1}^{m_{s_{1}s_{2}}%
\ast }(\overset{\symbol{94}}{\mathbf{r}}),
\end{equation}
\begin{equation}
\int d^{3}\mathbf{q}f^{J+1}(q)Y_{J+1}^{m_{s_{1}s_{2}}\ast }(\widehat{\mathbf{%
q}})e^{-i\mathbf{q}\cdot \mathbf{r}}=R_{n}^{J+1}(r)Y_{J+1}^{m_{s_{1}s_{2}}}(%
\overset{\symbol{94}}{\mathbf{r}}),
\end{equation}
where 
\begin{equation}
R_{n}^{\ell }\left( r\right) =-\frac{2}{n^{2}}\sqrt{\frac{\left( n-\ell
-1\right) !}{\left( \left( n+\ell \right) !\right) ^{3}}}e^{-r/n}\left( 
\frac{2r}{n}\right) ^{\ell }L_{n+\ell }^{2\ell +1}\left( \frac{2r}{n}\right)
.
\end{equation}
The associated Laguerre functions\ \ $L_{\lambda }^{\mu }\left( \rho \right) 
$\ \ are related to the confluent hypergeometric functions by 
\begin{equation}
L_{\lambda }^{\mu }\left( \rho \right) =\left( -1\right) ^{\mu }\frac{\left(
\lambda !\right) ^{2}}{\mu !\left( \lambda -\mu \right) !}F\left( -\lambda
+\mu ,\mu +1;\rho \right) .
\end{equation}
The generating function for the Laguerre functions is 
\begin{equation}
U_{\mu }\left( \rho ,u\right) \equiv \left( -1\right) ^{\mu }\frac{u^{\mu }}{%
\left( 1-u\right) ^{\mu +1}}\exp \left( -\frac{u\rho }{1-u}\right)
=\sum_{\lambda =\mu }^{\infty }\frac{L_{\lambda }^{\mu }\left( \rho \right) 
}{\lambda !}u^{\lambda },
\end{equation}
hence 
\begin{eqnarray}
&&\underset{\sigma _{1}\sigma _{2}s_{1}s_{2}}{\sum }C_{Jm_{J}12}^{s_{1}s_{2}%
\sigma _{1}\sigma _{2}}\int d^{3}\mathbf{p\,}d^{3}\mathbf{q\,}%
f^{J-1}(p)Y_{J-1}^{m_{s_{1}s_{2}}\ast }(\widehat{\mathbf{p}})\mathcal{M}%
_{s_{1}s_{2}\sigma _{1}\sigma _{2}}^{ope\left( s-s\right) }\left( \mathbf{p,q%
}\right) f^{J+1}(q)Y_{J+1}^{m_{\sigma _{1}\sigma _{2}}}(\widehat{\mathbf{q}})
\notag \\
&=&\underset{\sigma _{1}\sigma _{2}s_{1}s_{2}}{\sum }%
C_{Jm_{J}12}^{s_{1}s_{2}\sigma _{1}\sigma _{2}}\int d^{3}\mathbf{r\,}%
R_{n}^{J-1}\left( r\right) Y_{J-1}^{m_{s_{1}s_{2}}\ast }(\widehat{\mathbf{r}}%
)\left( 3\alpha \frac{\left( \overrightarrow{\sigma }_{2}\cdot \mathbf{r}%
\right) \left( \overrightarrow{\sigma }_{1}\cdot \mathbf{r}\right) }{16\pi
m_{1}m_{2}r^{5}}\right) R_{n}^{J+1}\left( r\right) Y_{J+1}^{m_{s_{1}s_{2}}}(%
\widehat{\mathbf{r}}) \\
&=&\frac{3\alpha }{16\pi m_{1}m_{2}}\int dr\,r^{2}\frac{1}{r^{3}}%
R_{n}^{J-1}\left( r\right) R_{n}^{J+1}\left( r\right) \times  \notag \\
&&\times \underset{\sigma _{1}\sigma _{2}s_{1}s_{2}}{\sum }%
C_{Jm_{J}12}^{s_{1}s_{2}\sigma _{1}\sigma _{2}}\int d\widehat{\mathbf{r}}%
\,Y_{J-1}^{m_{s_{1}s_{2}}\ast }(\widehat{\mathbf{r}})\left( \overrightarrow{%
\sigma }_{2}\cdot \widehat{\mathbf{r}}\right) \left( \overrightarrow{\sigma }%
_{1}\cdot \widehat{\mathbf{r}}\right) Y_{J+1}^{m_{s_{1}s_{2}}}(\widehat{%
\mathbf{r}}).  \notag
\end{eqnarray}
We can show, that

\begin{equation}
\underset{\sigma _{1}\sigma _{2}s_{1}s_{2}}{\sum }C_{Jm_{J}12}^{s_{1}s_{2}%
\sigma _{1}\sigma _{2}}\int d\widehat{\mathbf{r}}Y_{J-1}^{m_{s_{1}s_{2}}\ast
}(\widehat{\mathbf{r}})\left( \overrightarrow{\sigma }_{2}\cdot \widehat{%
\mathbf{r}}\right) \left( \overrightarrow{\sigma }_{1}\cdot \widehat{\mathbf{%
r}}\right) Y_{J+1}^{m_{s_{1}s_{2}}}(\widehat{\mathbf{r}})=\frac{1}{15}\frac{%
\sqrt{J\left( J+1\right) }}{2J+1},
\end{equation}
but 
\begin{equation}
\int_{0}^{\infty }dr\,r^{2}\frac{1}{r^{3}}R_{n}^{J-1}\left( r\right)
R_{n}^{J+1}\left( r\right) =0.
\end{equation}
The last expression can be proved in the following way. Let us consider the
more general case 
\begin{equation}
\int_{0}^{\infty }dr\,r^{\beta +2}R_{n}^{\ell }\left( r\right) R_{n}^{\ell
^{\prime }}\left( r\right) .
\end{equation}
The generating function for $R_{n}^{\ell }\left( r\right) $\ is

\begin{equation}
G_{n\ell }\left( r,u\right) =-\frac{2}{n^{2}}\sqrt{\frac{\left( n-\ell
-1\right) !}{\left( \left( n+\ell \right) !\right) ^{3}}}e^{-r/n}\left( 
\frac{2r}{n}\right) ^{\ell }\left( -1\right) ^{2\ell +1}\frac{u^{2\ell +1}}{%
\left( 1-u\right) ^{2\ell +2}}\exp \left\{ -\frac{u}{1-u}\frac{2r}{n}%
\right\} .
\end{equation}
Then we consider an expression 
\begin{eqnarray}
&&\int_{0}^{\infty }drr^{\beta +2}G_{n\ell }\left( r,u\right) G_{n\ell
^{\prime }}\left( r,v\right)  \notag \\
&=&\int_{0}^{\infty }drr^{\beta +2}\frac{4}{n^{4}}\sqrt{\frac{\left( n-\ell
-1\right) !\left( n-\ell ^{\prime }-1\right) !}{\left( \left( n+\ell \right)
!\right) ^{3}\left( \left( n+\ell ^{\prime }\right) !\right) ^{3}}}%
e^{-2r/n}\left( \frac{2r}{n}\right) ^{\ell +\ell ^{\prime }}\times  \notag \\
&&\times \frac{u^{2\ell +1}v^{2\ell ^{\prime }+1}}{\left( 1-u\right) ^{2\ell
+2}\left( 1-v\right) ^{2\ell ^{\prime }+2}}\exp \left\{ -\left( \frac{u}{1-u}%
+\frac{v}{1-v}\right) \frac{2r}{n}\right\} \\
&=&\frac{4}{n^{4}}\sqrt{\frac{\left( n-\ell -1\right) !\left( n-\ell
^{\prime }-1\right) !}{\left( \left( n+\ell \right) !\right) ^{3}\left(
\left( n+\ell ^{\prime }\right) !\right) ^{3}}}\frac{u^{2\ell +1}v^{2\ell
^{\prime }+1}}{\left( 1-u\right) ^{2\ell +2}\left( 1-v\right) ^{2\ell
^{\prime }+2}}\times  \notag \\
&&\times \int_{0}^{\infty }dr\left( \frac{2r}{n}\right) ^{\beta +2+\ell
+\ell ^{\prime }}\exp \left\{ -\left( 1+\frac{u}{1-u}+\frac{v}{1-v}\right) 
\frac{2r}{n}\right\} .  \notag
\end{eqnarray}
It is well known that 
\begin{equation}
\int_{0}^{\infty }d\rho \rho ^{\beta }e^{-\rho }=\Gamma \left( \beta
+1\right) .
\end{equation}
Hence 
\begin{eqnarray}
&&\int_{0}^{\infty }dr\left( \frac{2r}{n}\right) ^{\beta +2+\ell +\ell
^{\prime }}\exp \left\{ -\left( 1+\frac{u}{1-u}+\frac{v}{1-v}\right) \frac{2r%
}{n}\right\}  \notag \\
&=&\left( \frac{n}{2}\right) ^{\beta +3}\left( \frac{\left( 1-u\right)
\left( 1-v\right) }{1-uv}\right) ^{\beta +3+\ell +\ell ^{\prime }}\Gamma
\left( \beta +3+\ell +\ell ^{\prime }\right)
\end{eqnarray}
and 
\begin{eqnarray}
&&\int_{0}^{\infty }drr^{\beta +2}G_{n\ell }\left( r,u\right) G_{n\ell
^{\prime }}\left( r,v\right)  \notag \\
&=&\frac{2^{-\beta -1}}{n^{-\beta +1}}\sqrt{\frac{\left( n-\ell -1\right)
!\left( n-\ell ^{\prime }-1\right) !}{\left( \left( n+\ell \right) !\right)
^{3}\left( \left( n+\ell ^{\prime }\right) !\right) ^{3}}}\times \\
&&\times \frac{u^{2\ell +1}v^{2\ell ^{\prime }+1}\left( 1-u\right) ^{\beta
+1-\ell +\ell ^{\prime }}\left( 1-v\right) ^{\beta +1+\ell -\ell ^{\prime }}%
}{\left( 1-uv\right) ^{\beta +3+\ell +\ell ^{\prime }}}\Gamma \left( \beta
+3+\ell +\ell ^{\prime }\right) .  \notag
\end{eqnarray}
We expand this expression in a series: 
\begin{equation}
\int_{0}^{\infty }drr^{\beta +2}G_{n\ell }\left( r,u\right) G_{n\ell
^{\prime }}\left( r,v\right) =\sum_{\eta \eta ^{\prime }}C_{\eta \eta
^{\prime }}\left( n,\beta ,\ell ,\ell ^{\prime }\right) u^{\eta }u^{\eta
^{\prime }}.
\end{equation}
It is not difficult to show [31], that the coefficient $C_{n+\ell ,n+\ell
^{\prime }}$\ are given by the integral\ 
\begin{equation}
C_{n+\ell ,n+\ell ^{\prime }}\left( n,\beta ,\ell ,\ell ^{\prime }\right)
=\int_{0}^{\infty }drr^{\beta +2}R_{n}^{\ell }(r)R_{n}^{\ell ^{\prime }}(r).
\end{equation}
Simple but tedious calculations show that this\ coefficient is zero for $%
\beta =-3$, $\ell =J-1$, $\ell ^{\prime }=J+1$. Thus the kernel $\mathcal{K}%
_{12}$ does not contribute to the energy corrections at $O\left( \alpha
^{4}\right) $. The same result is obtained for the kernel $\mathcal{K}_{21}$.

\vskip0.8truecm {\normalsize \noindent {\textbf{\large References}}} 
{\normalsize \vskip0.4truecm }

{\normalsize \enumerate}

1. J. W. Darewych, Annales Fond. L. de Broglie (Paris) \textbf{23}, 15
(1998).

2. J. W. Darewych, in \textit{Causality and Locality in Modern Physics}, G
Hunter et al. (eds.), p. 333, (Kluwer 1998).

3. J. W. Darewych, Can. J. Phys. \textbf{76}, 523 (1998).

4. M. Barham and J. W. Darewych, J. Phys. A \textbf{31}, 3481 (1998).

5. B. Ding and J. Darewych, J. Phys. G \textbf{26}, 907 (2000).

6. J. D. Jackson, \textit{Classical Electrodynamics} (John Wiley, New York,
1975).

7. A. O. Barut, \textit{Electrodynamics and Classical Theory of Fields and
Particles} (Dover, New York, 1980).

8. W. T. Grandy, \textit{Relativistic Quantum Mechanics of Leptons and Fields%
} (Kluwer, 1991).

9. A. O. Barut, in \textit{Geometrical and Algebraic Aspects of Nonlinear
Field Theory, edited by} S. De Filippo, M. Marinaro, G. Marmo and G. Vilasi,
(Elsevier New York, 1989), p. 37.

10. A. G. Terekidi, J. W. Darewych, Preprint: arxiv, hep-th/0303250.

11. J. W. Darewych and L. Di Leo, J. Phys. A: Math. Gen. \textbf{29}, 6817
(1996).

12. J. W. Darewych and M. Horbatsch, J. Phys. B: At. Mol. Opt. \textbf{22},
973 (1989); \textbf{23}, 337 (1990).

13. W. Dykshoorn and R. Koniuk, Phys. Rev. A \textbf{41}, 64 (1990).

14. T. Zhang and R. Koniuk, Can. J. Phys. \textbf{70}, 683 (1992).

15. T. Zhang and R. Koniuk, Can. J. Phys. \textbf{70}, 670 (1992).

16. V. B. Berestetski, E. M. Lifshitz, and L. P. Pitaevski, \textit{Quantum
Electrodynamics }(Pergamon, New York, 1982).

17. G. Arfken and H. Weber \textit{Mathematical Methods for Physicists}
(Academic Press, 2001).

18. H. A. Bethe and E. E. Salpeter, \textit{Quantum Mechanics of One and
Two-Electron Atoms (Springer-Verlag/Academic, New York, 1957).}

19. M. A. Stroscio, Physics Reports C \textbf{22}, 5, 215 (1975).

20. John H. Connell, Phys. Rev. D \textbf{43}, 1393 (1991).

21. H. Hersbach, Phys. Rev. A \textbf{46}, 3657 (1992).

22. J. W. Darewych and A. Duriryak Phys. Rev. A \textbf{66}, 032102 (2002).

23. W. A. Barker, F. N. Glover, Phys. Rev. \textbf{99}, 317 (1955).

24. Vernon W. Hughes, Gisbert zu Putlitz in \textit{Quantum Electrodynamics
(World Scientific 1990), }T. Kinoshita, edit.

25. F. M. Pipkin in \textit{Quantum Electrodynamics (World Scientific 1990), 
}T. Kinoshita, edit.

26. H. Hellwig et al., IEEE Trans. Instrum. Meas. IM \textbf{19}, 200
(1970); L. Essen et al., Nature \textbf{229}, 110 (1971).

27. K. Pachucki, Phys. Rev. A \textbf{53}, 2092 (1996).

28. D. Bakalov, E. Milotti, C. Rizzo, A. Vacchi and E. Zavattini, Physics
Letters A \textbf{172}, 277 (1993)

29. M. I. Eides, H. Grotch, V. A. Shelyuto, Phys. Report \textbf{342} , 63
(2001).

30. J. R. Sapirstein, D. R. Yennie in \textit{Quantum Electrodynamics (World
Scientific 1990), }T. Kinoshita, edit.

31. M. Mizushima, \textit{Quantum Mechanics of Atomic Spectra and Atomic
Structure }(W. A. Benjamin, Inc., New York, 1970).

\end{document}